\documentclass[lettersize,journal]{IEEEtran}
\usepackage{amsmath,amsfonts}
\usepackage{amsthm}

\usepackage{algorithmic}
\usepackage{algorithm}
\usepackage{array}
\usepackage[caption=false,font=normalsize,labelfont=sf,textfont=sf]{subfig}
\usepackage{textcomp}
\usepackage{stfloats}
\usepackage{url}
\usepackage{verbatim}
\usepackage{graphicx}
\usepackage{cite}
\usepackage{tikz}
\usetikzlibrary{positioning}
\usepackage[table]{xcolor}
\definecolor{pfqnhighlight}{gray}{0.88}

\usepackage{xcolor}
\usepackage{booktabs}
\usepackage{multirow}
\usepackage{tikz}
\usepackage{pgfplots}
\usepgfplotslibrary{groupplots}
\pgfplotsset{compat=1.18}
\usetikzlibrary{arrows.meta,positioning,calc}
\usetikzlibrary{patterns.meta}

\definecolor{PFQNBlue}{RGB}{0,114,178}
\definecolor{PFQNCyan}{RGB}{230,159,0}
\definecolor{UROOrange}{RGB}{94,213,0}
\definecolor{PORed}{RGB}{213,94,0}
\definecolor{ITAGSGreen}{RGB}{0,158,115}
\definecolor{EFOPurple}{RGB}{204,121,167}
\usetikzlibrary{arrows.meta}
\usepackage[colorlinks=true,linkcolor=blue,citecolor=blue,urlcolor=blue]{hyperref}

\begin{document}

\title{Delay and Throughput Analysis of Computation Offloading in Mobile Edge Computing: A Queueing Network Approach}

\author{Amirparsa Bahrami and Farid Ashtiani,~\IEEEmembership{Senior Member,~IEEE}

\thanks{The authors are with the Department of Electrical Engineering, Sharif University of Technology, Tehran 14588-89694, Iran \newline(e-mail: amirparsa2828@gmail.com; ashtianimt@sharif.edu)}
\thanks{}}

\maketitle
\begin{abstract}
Mobile edge computing (MEC) enables mobile devices to offload
computation to nearby edge servers and to the cloud in order to
reduce end-to-end delay for different applications. In this paper, we study
static computation offloading when each task consists of multiple
dependent subtasks represented by a rooted directed tree. We develop
a product-form queueing-network (PFQN) model with an approximation to
capture the computation and communication dynamics of tree-structured
task execution in a multi-tier MEC system. Based on this model, we
derive closed-form expressions for server utilizations and
waiting times, and then construct a recursive algorithm for evaluating
the average delay of general tree-structured tasks. We formulate the
static offloading design problem as the minimization of the
rate-weighted average task delay over the routing probabilities, and
solve it through a differentiable optimization framework based on
softmax parameterization, log-sum-exp smoothing, and a stability
barrier on server utilizations. Numerical results show that the proposed PFQN approximation provides accurate delay estimates and that the delay-optimized static policy consistently outperforms the considered baseline algorithms in terms of both average task delay and maximum stable throughput.
\end{abstract}

\section{Introduction}
\label{sec:introduction}
\IEEEPARstart{L}OW latency is essential for many emerging mobile applications,
including augmented reality (AR) and virtual reality (VR), real-time inference, video analytics, and
sensor-driven analytics. In such systems, user experience depends
on end-to-end response time rather than only on raw computing
power~\cite{MaoMECsurvey,SatyanarayananEdge}. However, mobile
devices and Internet of Things (IoT) nodes are often limited in computation capability,
energy, and memory. This has made computation offloading a crucial
design mechanism: instead of processing every task locally, a device
may execute part or all of its workload on a nearby edge server or
in the cloud~\cite{DinhMCCsurvey,ChenMultiUserOffload16,DongSurvey2024}.

Mobile edge computing (MEC) provides the architectural support for
this paradigm by placing computing resources close to users while
still allowing access to remote cloud resources when needed
\cite{MaoMECsurvey,DinhMCCsurvey,SatyanarayananEdge}. In a
typical MEC system, the offloading decision affects multiple
interacting components of delay, including communication delay,
queueing delay, and computation delay. These components become
strongly coupled when multiple users share the same edge server or
cloud access link. Therefore, accurate delay modeling is important
for designing effective offloading policies.

A large body of prior work has studied computation offloading in MEC
from the viewpoints of delay minimization, energy efficiency,
resource allocation, robustness, and learning-based adaptation
\cite{ChenMultiUserOffload16,Sardellitti2015,QECO2025,DongSurvey2024,
WangRobustOffloadingTMC2023,QuRobustSchedulingTMC2022,
LiResilienceJSAC2023,JiaoJCC24,AleTCCN21}.

 In particular, queueing-theoretic
analysis has long been used in MEC to capture congestion effects and
derive latency-aware offloading rules. Existing queueing-based MEC
studies, however, mainly fall into three groups.

First, several works use relatively simple queueing models to
represent one bottleneck or one stage of the offloading path, and
then optimize task assignment or server selection on top of that
model. For example, Xue \emph{et al.} study task allocation in a
5G heterogeneous MEC network using a queueing-theoretic model,
while Katayama and Tachibana derive delay expressions for a MEC
platform with a dedicated MEC server, a shared MEC server, and a
cloud server, where the shared server is modeled as a queueing
bottleneck \cite{XueQueueMEC,KatayamaQueueSensors22}. These works
demonstrate that queueing models are useful for latency-aware
offloading, but they mainly treat tasks as monolithic jobs and focus
on server-level congestion rather than the internal traffic
dependencies created by structured applications.

Second, another line of work uses queueing models to derive
delay-related or quality-of-service (QoS)-oriented performance measures in multi-tier
offloading systems. For example, queue-length-based offloading
rules have been proposed for delay-sensitive applications in
federated cloud-edge-fog systems by explicitly targeting the
probability of QoS violation, and Markovian queueing models have
been used for end-to-end delay analysis in computation offloading
systems \cite{BairagiQueueLengthCCNC24,XieMarkov23}. These works are
analytically valuable, but the traffic they study is still generated
by single-stage tasks or independent requests, rather than by
structured task execution in which one completed component may
trigger multiple downstream operations.

Third, some recent MEC studies use queue or workload information,
often together with learning-based control, to better capture
congestion and dynamic system behavior. Recent works include
quality-of-experience (QoE)-oriented deep reinforcement learning for distributed MEC
offloading~\cite{QECO2025}, as well as more recent dependency-aware
deep reinforcement learning approaches that explicitly consider task
graphs in MEC environments~\cite{DependentTaskGraphDRL2025,Zhai2025Joint}.
These approaches focus on learning-based offloading decisions.
In particular,~\cite{QECO2025} uses queue information but does not model
dependencies among subtasks. 

On the other hand, modern MEC applications are rarely monolithic.
Instead, they often consist of multiple dependent subtasks, where
the output of one subtask is needed before another one can start. Task graphs provide a natural representation for such workloads, and several MEC and edge-computing studies have used directed acyclic graph (DAG)-based models to express subtask precedence, placement, and offloading decisions \cite{TaskGraphMEC,LiIBDASH22,SundarLiang2018,
ShuZhaoHanMinDuan2019,CaoDependentOffloading2023,
DependentTaskGraphDRL2025}.

 In this paper, we focus on the important
special case in which the task graph is a rooted directed tree. This
structure still captures branching and multi-stage applications,
while avoiding some of the combinatorial complexity of general
DAGs.

The key difficulty is that task dependencies fundamentally change
the traffic seen by the servers. Even if external task arrivals are
Poisson, the completion of one subtask may immediately trigger a
batch of computation load units or data units for its children.
Hence, the internal traffic is correlated and bursty rather than
independent. Existing queueing-based MEC works typically model congestion caused
by multiple users sharing computing servers or communication
bottlenecks, and they use these models to derive delay-aware
offloading rules. However, to the best of our knowledge, they do not
explicitly capture the additional correlation created when the
completion of one subtask initiates a batch of downstream computations
or communication operations. This is the main gap addressed in this
work.

To bridge this gap, we develop an analytical framework for static
computation offloading in multi-tier MEC systems with
tree-structured task graphs. Our approach models the system through
a product-form queueing-network (PFQN) approximation with a
congestion-correction step that preserves tractability while
capturing the additional delay induced by correlated batched
internal traffic. Using the resulting closed-form utilizations and
waiting times, we derive a recursive delay-evaluation algorithm for
general task trees and optimize the routing probabilities of a
static offloading policy through a differentiable objective function.

Compared with recent learning-based MEC offloading works, our
framework provides an interpretable analytical surrogate for delay
evaluation and policy optimization. Compared with existing
queueing-based MEC studies, it explicitly links dependent subtask
execution, the correlated internal traffic created by task
progression, and end-to-end task delay under a static routing
policy. This makes the framework particularly suitable for fast
performance evaluation and static policy design in structured MEC
workloads.

The main contributions of this work are summarized as follows:
\begin{itemize}
\item We develop a PFQN-based analytical framework for multi-tier
MEC systems with tree-structured task graphs, including a
congestion-correction mechanism that captures the delay effect of
correlated batched internal traffic and yields closed-form
expressions for effective utilizations and waiting times.

\item We derive a recursive algorithm for evaluating the average
delay of general tree-structured tasks and formulate the static
offloading problem as the minimization of the rate-weighted
average task delay over the routing probabilities.

\item We solve the resulting optimization problem through a
differentiable framework based on softmax parameterization,
log-sum-exp smoothing, and a stability barrier on server
utilizations.

\item Through numerical experiments, we show that the proposed approximation remains accurate over a broad set of stable routing policies, and that the delay-optimized static policy consistently outperforms the considered baseline algorithms in both average task delay and maximum stable throughput.
\end{itemize}

The rest of the paper is organized as follows. Section~\ref{sec:system-model}
presents the system and task models. Section~\ref{sec:qn_model}
develops the PFQN-based queueing approximation and the
congestion-correction step. Section~\ref{sec:delay-derivation}
derives the average task-delay metric and presents the recursive
delay-evaluation algorithm. Section~\ref{sec:optimization}
formulates and solves the static offloading optimization problem.
Section~\ref{subsec:numerical} reports numerical validation and
compares the optimized policy with baseline algorithms.

Table~\ref{tab:notation} summarizes the main notation used throughout the paper.

\begin{table*}[!t]
\caption{Summary of Main Notation}
\label{tab:notation}
\centering
\footnotesize
\setlength{\tabcolsep}{4pt}
\renewcommand{\arraystretch}{1.12}
\begin{tabular}{
    p{0.105\textwidth}
    p{0.355\textwidth}
    p{0.105\textwidth}
    p{0.355\textwidth}
}
\hline
\textbf{Symbol} & \textbf{Description}
& \textbf{Symbol} & \textbf{Description} \\
\hline

$N$ &
Total number of computing and communication servers
&
$S_n$ &
Server indexed by $n$
\\

$N_{\mathrm{user}}$, $N_{\mathrm{edge}}$, $N_{\mathrm{cloud}}$ &
Numbers of user devices, MEC servers, and cloud servers
&
$\mathcal{E}$, $\mathcal{C}$ &
Sets of edge and cloud servers, respectively
\\

$Z_n$ &
Connection zone of server $S_n$
&
$\mu_n$, $\mu'_n$ &
Service rate of edge server $S_n$ and transmission rate of cloud queue $S_n$
\\
${T}_k$ &
Task of type $k$
&
$M_k$, $v_{km}$ &
Number of subtasks in ${T}_k$ and its $m$th subtask
\\

$\mathcal{C}(v_{km})$ &
Set of children of subtask $v_{km}$
&
$\mathrm{FC}(v_{km})$ &
Child of $v_{km}$ having the smallest BFS index
\\

$l_{km}$ &
Number of computation load units of subtask $v_{km}$
&
$r_{km}$ &
Number of result-data units produced by subtask $v_{km}$
\\

$d_{(ki,kj)}$ &
Dependency-data units transferred from $v_{ki}$ to $v_{kj}$
&
$d_{(\mathrm{out},k1)}$ &
Input-data units required to initiate the root subtask
\\

$\lambda_k^{(n)}$ &
External arrival rate of type-$k$ tasks generated by user $S_n$
&
$\lambda_k$ &
Aggregate external arrival rate of type-$k$ tasks

\\

$R_{k1}^{(n_0,n)}$ &
Probability of routing the root of a task generated at $S_{n_0}$ to $S_n$
&
$R_{ki}^{(n_p,n)}$ &
Probability of routing child $v_{ki}$ from its parent server $S_{n_p}$ to $S_n$
\\

$R$ &
Collection of all static routing probabilities
&
$D_{km,n}^{\mathrm{cloud}}$ &
Aggregated data workload when $v_{km}$ and its descendants are assigned to cloud server $S_n$
\\

$B_{km,n}$ &
Batch size generated when subtask $v_{km}$ is assigned to $S_n$
&
$\gamma_{km,n}(R)$ &
Initiation rate of subtask $v_{km}$ at server $S_n$
\\

$\Lambda_n(R)$ &
Nominal arrival rate of individual computation or data units at $S_n$
&
$C_n(R)$ &
Batch-correlation congestion-correction term at $S_n$
\\

$\alpha_n(R)$ &
Rate of the redundant Poisson stream added at $S_n$
&
$\rho_n(R)$ &
Batch-corrected utilization of server $S_n$
\\

$W_n(R)$ &
Average queueing waiting time at server $S_n$
&
$(a_2,\ldots,a_D)$ &
Execution-server scenario for the children of a star-shaped task
\\

$P_n(a)$ &
Probability of scenario $a$ conditioned on completion of the parent at $S_n$
&
$x_j^{(ki)}$ &
Mean processing or transmission time of child $v_{ki}$ at server $S_j$
\\

$T_j(a)$ &
Delay contribution of server $S_j$ under scenario $a$
&
$ET_{km}^{(n)}(R)$ &
Recursively evaluated delay of the subtree rooted at $v_{km}$ when its root is processed at $S_n$
\\

$\operatorname{TaskET}({T}_k,R)$ &
Average completion delay of task type $k$
&
$T_{\mathrm{tot}}(R)$ &
Network-wide rate-weighted average task delay
\\

\hline
\end{tabular}
\end{table*}

\section{System Model}
\label{sec:system-model}

In this section, we describe the communication--computation
infrastructure, the task structure, the offloading mechanism,
and the definition of task completion delay. 

\subsection{Computing Nodes}
\label{subsec:computing-nodes}

We consider a heterogeneous computing system consisting of
\(N\) servers, indexed by
\[
  \{S_1, S_2, \ldots, S_N\},
\]
which may represent mobile devices, mobile edge computing
(MEC) servers, or mobile cloud computing (MCC) servers
\cite{MaoMECsurvey,DinhMCCsurvey}. We denote by
\(N_{\mathrm{user}}\), \(N_{\mathrm{edge}}\), and
\(N_{\mathrm{cloud}}\) the numbers of mobile users, edge servers,
and cloud servers, respectively, so that
\[
  N = N_{\mathrm{user}} + N_{\mathrm{edge}} + N_{\mathrm{cloud}}.
\]

We explicitly distinguish between servers located at the edge of the
network (mobile devices and MEC servers) and servers hosted in remote
data centers, because these two categories have markedly different
latency and computational characteristics; this edge--cloud hierarchy
is standard in MEC and mobile cloud computing architectures
\cite{SatyanarayananEdge,MaoMECsurvey}.

The \emph{edge set} contains all mobile users and MEC
servers, i.e.,
\begin{equation}
  \mathcal{E} =
  \big\{ S_n \,\big|\,
  1 \le n \le N_{\mathrm{user}} + N_{\mathrm{edge}} \big\},
  \label{eq:setE}
\end{equation}
while the \emph{cloud set} contains all MCC servers, i.e.,
\begin{equation}
  \mathcal{C} =
  \big\{ S_n \,\big|\,
  N - N_{\mathrm{cloud}} < n \le N \big\}.
  \label{eq:setC}
\end{equation}

Each server \(S_n\) is modeled as a single-server
first come first served (FCFS) queue with exponential service
time. The mean service time at \(S_n\) is \(1/\mu_n\), where
\(\mu_n\) is the (computational or transmission) service rate.
This M/M/1-type abstraction is widely used in queueing-theoretic
studies of MEC and task offloading
\cite{XueQueueMEC,MaoMECsurvey}. Although the framework can be extended
to more general service time distributions at the cost of increased analytical and
computational complexity.

Servers may offload subtasks to other servers within their
\emph{connection zone}. The connection zone of server \(S_n\)
is denoted by
\begin{equation}
  \mathcal{Z}_n
  =
  \big\{ S_i \,\big|\,
  S_i \text{ is in the connection zone of } S_n \big\},
  \label{eq:Zn}
\end{equation}
and we always have \(S_n \in \mathcal{Z}_n\) for
\(1 \le n \le N\). This connectivity-constrained offloading model,
where each node can only forward tasks to a subset of servers, also
appeared in recent MEC task-allocation formulations
\cite{XueQueueMEC}.

\subsection{Task Model}
\label{subsec:task-model}

Each mobile user generates computation tasks that can be
decomposed into a set of dependent subtasks. We explicitly
model the dependencies among subtasks using a directed
rooted tree, as is standard in task-graph based offloading
for MEC and mobile cloud systems~\cite{TaskGraphMEC}.

There are \(K\) task types in the system. A task of type \(k\)
is denoted by \(T_k\) and represented as
\[
  T_k = \big( V(T_k), E(T_k) \big),
\]
where \(V(T_k)\) is the set of vertices (subtasks) and
\(E(T_k)\) is the set of directed edges (precedence constraints).
We write
\begin{equation}
  V(T_k) = \{ v_{k1}, v_{k2}, \ldots, v_{kM_k} \},
  \label{eq:vertices}
\end{equation}
where \(M_k\) is the number of subtasks in type-\(k\) tasks.
The directed edges encode data dependencies:
\begin{equation}
  E(T_k)
  =
  \big\{ (v_{ki}, v_{kj}) \,\big|\,
  \text{there is an edge from } v_{ki}
  \text{ to } v_{kj} \big\}.
  \label{eq:edges}
\end{equation}
An edge \((v_{ki}, v_{kj})\) indicates that part of the output
produced by subtask \(v_{ki}\) is needed as input to subtask
\(v_{kj}\), and therefore \(v_{kj}\) can start only after \(v_{ki}\)
has completed and its result is available at the server where
\(v_{kj}\) is executed.

Vertices are indexed according to a breadth-first search
(BFS) traversal starting from the root of the tree. Specifically,
we scan the tree level by level, from the root to the leaves, and
assign indices in the order in which the subtasks are
visited~\cite{CormenAlgorithms09}.
\begin{figure}[t]
  \centering
  \begin{tikzpicture}[
    level distance=1.3cm,
    sibling distance=2.2cm,
    every node/.style={circle,draw,minimum size=7mm,inner sep=0pt,font=\footnotesize},
    level 1/.style={sibling distance=2.2cm},
    level 2/.style={sibling distance=1.6cm}
  ]
    \node (v1) {$v_{k1}$}
      child { node (v2) {$v_{k2}$}
        child { node (v4) {$v_{k4}$} }
        child { node (v5) {$v_{k5}$} }
      }
      child { node (v3) {$v_{k3}$}
        child { node (v6) {$v_{k6}$} }
      };
  \end{tikzpicture}
  \caption{Example of BFS indexing for a task tree.}
  \label{fig:bfs-indexing}
\end{figure}
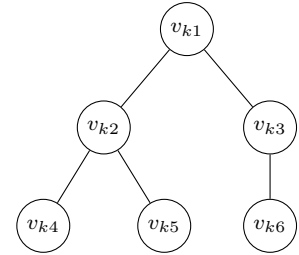

For each subtask \(v_{km}\) of task \(T_k\), we denote by
\begin{equation}
  \mathcal{C}(v_{km})
  =
  \big\{ v_{ki} \,\big|\,
  (v_{km}, v_{ki}) \in E(T_k) \big\}
  \label{eq:children-set}
\end{equation}
its set of children in the task tree. When needed, we refer to
the child in \(\mathcal{C}(v_{km})\) with the smallest BFS
index as the first child and denote it by \(\mathrm{FC}(v_{km})\).

\medskip
\noindent\textbf{Computation load units.}
We model the computation associated with each subtask
at a finer granularity by introducing the notion of a
\emph{computation load unit}. A load unit is the smallest
amount of computation that can be processed by a
server. A larger or more complex subtask is modeled as the
aggregation of multiple load units, following standard
queueing-theoretic modeling of phase-type service
times~\cite{KleinrockQueueing75}.

For the sake of simplicity, the computation load of subtask \(v_{km}\) is
described by a positive integer \(l_{km}\), meaning that
computing \(v_{km}\) consists of processing \(l_{km}\) independent
load units. When the subtask is executed at server \(S_n\),
each load unit experiences an exponential service time with
mean \(1/\mu_n\). Hence, the total computation time of
\(v_{km}\) at \(S_n\) is the sum of \(l_{km}\) i.i.d.\ exponential
random variables, i.e., an Erlang distribution~\cite{KleinrockQueueing75}.
More general distributions could also be incorporated into our model, at the cost of more complexity.

\medskip
\noindent\textbf{Dependency and result data units.}
We model communication at a similar level of granularity.
A \emph{data unit} is the smallest indivisible amount of data
that can be transmitted over a link.

For each edge \((v_{ki}, v_{kj}) \in E(T_k)\), we let
\(d_{(ki,kj)}\) denote the number of \emph{dependency data
units} that must be transferred from the server executing
\(v_{ki}\) to the server executing \(v_{kj}\) so that \(v_{kj}\) can
start. These dependency data units are formed from (part of)
the result produced by subtask \(v_{ki}\) and contain the
information needed to initiate the computation of
\(v_{kj}\). We also define \(d_{(\mathrm{out},k1)}\) as the number of data
units that must be transmitted from the originating mobile
user to the server that executes the root subtask \(v_{k1}\),
in order to provide the input data required to initiate \(v_{k1}\).

Each subtask produces a certain amount of result data that
must eventually be delivered back to the task origin when the
task completes. We denote by \(r_{km}\) the number of result
data units associated with subtask \(v_{km}\). If \(v_{km}\) is not
a leaf in the task tree, we set \(r_{km} = 0\).

The time required to transmit a single data unit over a given
communication link is modeled as an exponential random variable.
This memoryless approximation captures the
service time variability introduced by wireless channel fluctuations,
packet errors, and possible
retransmissions~\cite{ZhangWirelessService23}. Hence, the time to
transmit \(d\) data units over that link is the sum of \(d\)
independent exponential random variables, i.e., an Erlang
distribution, which is consistent with classical queueing
models~\cite{KleinrockQueueing75}.

\medskip
\noindent\textbf{Illustrative example.}
Consider a parent subtask \(v_{k1}\) that completes at server
\(S_1\) and has two children, \(v_{k2}\) and \(v_{k3}\), with
\(d_{(k1,k2)}=3\) and \(d_{(k1,k3)}=2\). Suppose that \(v_{k2}\)
is assigned to server \(S_2\), while \(v_{k3}\) is assigned to
the same server \(S_1\) as its parent. In this case, the three
dependency data units associated with the edge
\((v_{k1},v_{k2})\) are transmitted from \(S_1\) to \(S_2\),
whereas no data transmission is required for the edge
\((v_{k1},v_{k3})\). Thus, routing decisions are made at the
subtask level: all dependency data units associated with an edge
follow the server assignment of the corresponding child subtask
and are not routed independently.

\subsection{Offloading Model and Task Completion Time}
\label{subsec:offloading-model}

\medskip
\noindent\textbf{Task arrivals.}
Mobile user \(S_n\), \(1 \le n \le N_{\mathrm{user}}\), generates tasks
of type \(k \in \{1,\ldots,K\}\) according to a Poisson process
with rate \(\lambda_k^{(n)}\). Different users and task types are
independent, which is a standard assumption in queueing-based
models of MEC and mobile cloud offloading~\cite{ChenMultiUserOffload16,MaoMECsurvey}.

\medskip
\noindent\textbf{Execution of subtasks.}
When a new type-\(k\) task is generated at user \(S_n\), its root
subtask \(v_{k1}\) can either be processed locally at \(S_n\) or
offloaded to some server \(S_j \in \mathcal{Z}_n\). Offloading
\(v_{k1}\) incurs a transmission delay that depends on the
number of input data units \(d_{(\mathrm{out},k1)}\) and the
communication rate between \(S_n\) and \(S_j\).

More generally, suppose subtask \(v_{km}\) is executed at server \(S_n\).
For each child \(v_{ki} \in \mathcal{C}(v_{km})\) there are two options:
(i) compute \(v_{ki}\) at the same server \(S_n\), or
(ii) offload \(v_{ki}\) to some \(S_j \in \mathcal{Z}_n\).
If \(v_{ki}\) is offloaded from \(S_n\) to \(S_j\), the dependency
data consisting of \(d_{(km,ki)}\) units (derived from the
result of \(v_{km}\)) must be transmitted from \(S_n\) to \(S_j\),
incurring an additional transmission delay. If \(v_{ki}\) is
processed locally at \(S_n\), no dependency data transmission delay
arises for that edge.

We denote by \(\mathrm{start}_{km}\) and \(\mathrm{end}_{km}\) the
time instants at which processing of \(v_{km}\) starts and
completes (including both communication and computation
for that subtask), respectively.

\medskip
\noindent\textbf{Task completion delay.}
The completion time of an entire type-\(k\) task is defined as
the time elapsed between the start of the first subtask and
the completion of the last subtask:
\begin{equation}
  D_k
  =
  \max_{1 \le m \le M_k} \mathrm{end}_{km}
  -
  \min_{1 \le m \le M_k} \mathrm{start}_{km}.
  \label{eq:task-delay-general}
\end{equation}
Since the task graph is a rooted tree, the earliest-starting
subtask is always the root \(v_{k1}\). Hence
\begin{equation}
  D_k
  =
  \max_{1 \le m \le M_k} \mathrm{end}_{km}
  -
  \mathrm{start}_{k1}.
  \label{eq:task-delay-root}
\end{equation}

\medskip
\noindent\textbf{Edge--cloud tradeoffs.}
At the edge of the network (mobile users and MEC
servers), the main bottleneck is the limited computational
power, while communication between nearby nodes is
relatively fast. Because of this, we assume that, for
intra-edge offloading (e.g., between a user and an MEC
server or between two MEC servers), the transmission delays of dependency and result
data are negligible compared to the computation delays.
In other words, for offloading decisions that remain
inside \(\mathcal{E}\), the dominant contribution to delay is
computation, in line with standard MEC system models~\cite{MaoMECsurvey}.

Cloud servers in $\mathcal{C}$ reside in remote data centers.
They typically offer much larger computational capacity than
edge servers, but communication to and from them incurs higher
latency than local communication. To capture this tradeoff, we
model computation at cloud servers as effectively instantaneous
(we take $\mu_n = \infty$ for $S_n \in \mathcal{C}$), while
communication between the edge and each cloud server
$S_n \in \mathcal{C}$ is represented by a dedicated FCFS
queue (the \emph{cloud queue}) with exponential service rate
$\mu_n'$, which corresponds to the available data rate on
the edge--cloud link~\cite{DinhMCCsurvey,MaoMECsurvey}.

Thus, when a subtask (or a group of subtasks) is offloaded
to cloud server \(S_n\), we only model the time spent in the
cloud queue of \(S_n\); once the required data have reached
the cloud, the computation is assumed to complete
effectively instantaneously.

A key modeling assumption is that, once a subtask is
offloaded to the cloud, all of its descendant subtasks in the
task tree are also executed in the \emph{same} cloud server.
This reflects the following practical consideration: after
paying the cost of sending the intermediate result from the
edge to the cloud, it is usually preferable to keep subsequent
processing in the cloud, where computational resources are
abundant; moreover, moving intermediate results repeatedly
between edge and cloud would incur additional end-to-end
delay and extra data transfer overhead, which is typically
undesirable for latency sensitive applications~\cite{TaskGraphMEC}.

Under this assumption, when subtask \(v_{km}\) is offloaded
to cloud server \(S_n \in \mathcal{C}\), we aggregate the dependency data needed to
start \(v_{km}\) (either from the task origin if \(m = 1\), or from
its parent subtask if \(m \ge 2\)) together with all result data
produced by the descendants of \(v_{km}\), which must
eventually be returned to the edge when the task completes, into a
single cloud-queue workload.

Formally, we define the total number of data units injected
into the cloud queue of \(S_n\) as
\begin{equation}
  \begin{aligned}
  D^{\mathrm{cloud}}_{km,n}
  &=
  d_{(\mathrm{out},km)} \, \mathbf{1}(m = 1) + r_{km}
  \\
  &\quad
  + d_{(kj,km)} \, \mathbf{1}\big(2 \le m \le M_k,\,
     v_{km} \in \mathcal{C}(v_{kj})\big)
  \\
  &\quad
  + \sum_{v_{ki} \in \text{descendants of } v_{km}} r_{ki},
  \end{aligned}
  \label{eq:DRkm}
\end{equation}
where \(d_{(\mathrm{out},km)}\) and \(d_{(kj,km)}\) are defined
as in the previous subsection, \(\mathbf{1}(\cdot)\) is the
indicator function, and the summation is taken over all
descendants \(v_{ki}\) of \(v_{km}\) in the task tree.

\subsection{Static Computation Offloading Policy}
\label{subsec:static-offloading}

Given the task and system models above, the main design
degree of freedom is \emph{where} each subtask is executed.
We focus on \emph{static} computation offloading policies:
offloading decisions are expressed as fixed routing
probabilities that do not depend on instantaneous queue
states.

In general, two broad classes of offloading policies can be distinguished:
\emph{static policies}, in which the server–allocation decisions for
subtasks are specified by fixed probabilities and do not depend on
instantaneous queue lengths or other fast-varying system state; and
\emph{dynamic policies}, in which offloading decisions are updated at
each decision epoch based on the current state of the system (e.g.,
queue lengths, estimated delays). This classification is widely used
in the MEC offloading literature~\cite{MaoMECsurvey,ChenMultiUserOffload16}.

Dynamic policies can in principle react to congestion in
real time, but they incur significant signaling and
coordination overhead, especially when the number of
servers and users is large. In contrast, static policies only
require statistical information about arrival rates and service
capabilities, and the offloading decisions can be precomputed
and stored as routing tables at the servers. In this work,
we adopt a static policy and optimize its routing
probabilities to minimize the average task completion delay.

\medskip
\noindent\textbf{Routing of root subtasks.}
Consider a task of type \(k \in \{1,\ldots,K\}\) generated by
mobile user \(S_{n_0}\), where \(1 \le n_0 \le N_{\mathrm{user}}\).
The root subtask of this task is \(v_{k1}\). When the task is
generated, the system decides whether \(v_{k1}\) is
processed locally at \(S_{n_0}\) or offloaded to some server
\(S_n\) in the connection zone \(\mathcal{Z}_{n_0}\).

We denote by
\begin{equation}
  R_{k1}^{(n_0,n)}
\end{equation}
the probability that the root subtask \(v_{k1}\) of a type-\(k\)
task generated at user \(S_{n_0}\) is sent for execution to
server \(S_n\). These root-routing probabilities satisfy:
\begin{align}
  &\sum_{n \in \mathcal{Z}_{n_0}} R_{k1}^{(n_0,n)} = 1,
  && 1 \le k \le K,\ 1 \le n_0 \le N_{\mathrm{user}},
  \label{eq:root-normalization}
  \\
  &R_{k1}^{(n_0,n)} = 0,
  && n \notin \mathcal{Z}_{n_0}.
  \label{eq:root-connection-zone}
\end{align}
Hence, the root of each arriving task is always sent to one
of the servers in the user's connection zone, according to a
fixed probability distribution.

\medskip
\noindent\textbf{Routing of child subtasks.}
Now consider a generic subtask \(v_{km}\) of a type-\(k\) task.
For each child \(v_{ki} \in \mathcal{C}(v_{km})\), we define
the routing probabilities
\begin{equation}
  R_{ki}^{(n_p,n)}, \qquad
  S_n \in \mathcal{Z}_{n_p},
\end{equation}
as the probability that child subtask \(v_{ki}\) is assigned to
server \(S_n\) when its parent subtask \(v_{km}\) has just
completed at server \(S_{n_p}\). These probabilities satisfy
\begin{align}
  &\sum_{S_n \in \mathcal{Z}_{n_p}}
  R_{ki}^{(n_p,n)} = 1,
  \label{eq:child-normalization}
  \\
  &R_{ki}^{(n_p,n)} = 0,
  && S_n \notin \mathcal{Z}_{n_p},
  \label{eq:child-connection}
\end{align}
for all task types \(k\), subtask indices \(m\) and \(i\) with
\(v_{ki} \in \mathcal{C}(v_{km})\), and all possible
parent-completing servers \(S_{n_p} \in \mathcal{E}\).

Thus, for every edge \((v_{km},v_{ki})\) in the task tree and
every possible parent-completing server \(S_{n_p}\), the static
policy specifies a probability distribution over the servers in
\(\mathcal{Z}_{n_p}\) that may execute \(v_{ki}\).

\medskip
\noindent\textbf{Policy representation.}
The static computation offloading policy of the whole
network is the collection of all routing probability variables:
\begin{equation}
  R =
  \big\{
    R_{k1}^{(n_0,n)},
    R_{ki}^{(n_p,n)}
    \,\big|\,
    \text{all feasible } k,m,i,n_0,n_p,n
  \big\}.
\end{equation}
Here, \(n_0\) denotes the index of the user that generates the
task, whereas \(n_p\) denotes the index of the server at which
the parent of a non-root subtask completes. For root-routing
variables, feasibility requires
\(S_n \in \mathcal{Z}_{n_0}\). For child-routing variables,
feasibility requires \(S_{n_p} \in \mathcal{E}\),
\(S_n \in \mathcal{Z}_{n_p}\), and
\(v_{ki} \in \mathcal{C}(v_{km})\). Routing probabilities
corresponding to destinations outside the relevant connection
zone are fixed to zero and are not treated as decision variables.

\section{Queueing-Network Model for Computation Offloading}
\label{sec:qn_model}
In this section, we develop a product-form queueing-network
(PFQN) approximation for the computation offloading model of
Section~\ref{sec:system-model}. Here, PFQN refers to a queueing
network with a product-form steady-state distribution, which
allows the network to be analyzed through per-server effective
arrival rates, utilizations, and waiting times. The main
difficulty is that when a subtask is initiated at a server, it
does not contribute a single unit to the corresponding queue.
Rather, it creates a \emph{batch} of units: either several
computation load units at an edge server or several data units
at a cloud queue. Therefore, the traffic generated by task
execution is correlated and appears in back-to-back bursts.
This effect must be captured in the queueing model in order to
obtain accurate waiting-time expressions.
In the classical product-form setting, the network nodes must
satisfy quasi-reversibility conditions, which in particular
require that the effective traffic exchanged among queues be
compatible with Poisson-type flows
\cite{BCMP75,Kelly79,Serfozo99,Gelenbe93}. The original
offloading system studied here does not satisfy this condition
directly. Indeed, when a subtask is initiated at a server, it
creates a batch of computation load units or data units that
arrive back-to-back at the corresponding queue. Hence, the
internal traffic is correlated and is not naturally of the form
required by standard product-form queueing-network results.

For this reason, the PFQN used in this paper should be
understood as an approximation rather than an exact
representation of the original system. The purpose of the
approximation is to replace each correlated batch by an
equivalent Poisson flow, i.e., a Poisson flow with a suitable rate. After this transformation, each server
is modeled as a quasi-reversible M/M/1 node, and the resulting
approximating network admits a product-form interpretation.

\subsection{From Single-Queue Approximation to a Network of Queues}

In~\cite{RahnamaniaAshtianiCorrelatedArrivals}, a
redundant-arrival approximation was developed for a single
M/M/1 queue with batches of size two in order to preserve the average queueing delay. In~\cite{RahnamaniaAshtianiTriggeredRequests} a more exact approximation was proposed. However, it cannot be extended to a network easily.
In this work, we
generalize \cite{RahnamaniaAshtianiCorrelatedArrivals} to arbitrary deterministic batch
sizes and then extend it to a network of queues. Consider an M/M/1 queue fed by Poisson arrivals of batches.
Suppose that a batch of type $b$ arrives with rate $\lambda_b$ and contains
exactly $L_b$ customers. In the original queue, the first customer of the
batch arrives as an ordinary Poisson arrival and therefore sees the queue in
steady state according to Poisson arrivals see time averages (PASTA) property~\cite{KleinrockQueueing75}. However, the remaining $L_b-1$ customers arrive immediately
after the first one and thus do not see the queue in steady state; instead,
each of them sees at least the customers that arrived before it in the same
batch, and hence experiences more congestion than an independent Poisson
arrival with the same average rate.
 Motivated by the size-two construction in
\cite{RahnamaniaAshtianiCorrelatedArrivals}, our generalized
approximation replaces an arbitrary batch of size \(L_b\) by
a dual quasi-reversible M/M/1 queue with two ingredients:
\begin{enumerate}
    \item a Poisson stream of individual customers with rate $L_b \lambda_b$, which represents the same average offered load;
    \item an additional independent Poisson stream of redundant customers, whose role is only to reproduce the extra congestion caused by the within-batch correlation.
\end{enumerate}
Because the resulting queue is quasi-reversible, its departures are Poisson, and the same idea can be used again if the output of one queue starts a new batch somewhere else.

Having generalized the single queue construction to arbitrary
batch sizes, we further extend it from a single queue to a
network of queues. Whenever a subtask initiates a batch at some queue, we first replace that batch by an equivalent Poisson stream of individual units with the same average rate. Then we add an extra independent Poisson stream whose rate is chosen so as to recover the congestion ignored by this Poissonization. Since the completion of one batch may initiate another batch at the same queue or at another queue, the same construction is applied repeatedly throughout the task graph. In this way, the entire computation offloading system is approximated by a network of quasi-reversible M/M/1 queues.

\subsection{Batch Types Created by Subtask Execution}

For every task type $T_k$ and every subtask $v_{km}$, define the batch size created when $v_{km}$ is assigned to server $S_n$ as
\begin{equation}
B_{km,n} =
\begin{cases}
l_{km}, & S_n \in \mathcal{E},\\[1mm]
D^{\mathrm{cloud}}_{km,n}, & S_n \in \mathcal{C}.
\end{cases}
\label{eq:Bkmn}
\end{equation}
Thus, if $v_{km}$ is processed at an edge server, the corresponding batch consists of its $l_{km}$ computation load units; if it is offloaded to a cloud server, the corresponding batch consists of the aggregated number of data units that must traverse the cloud queue.

The approximation keeps track of the \emph{rate at which such batches are initiated}. Once these batch-initiation rates are known, the queueing approximation of each server follows directly.

\subsection{Traffic Equations for Batch Initiations}

We now derive the traffic equations. Since the task graph is a rooted tree, every subtask has exactly one parent except the root. Therefore, the rate at which a subtask is initiated can be computed recursively from the initiation rate of its parent.

Let
\[
\gamma_{km,n}(R)
\]
denote the steady-state rate at which subtask $v_{km}$ is initiated at server $S_n$ under routing policy $R$. Equivalently, this is the rate of the first unit of the batch associated with $v_{km}$ at server $S_n$.

For the root subtask $v_{k1}$, the initiation rate at server $S_n$ is
\begin{equation}
\gamma_{k1,n}(R)
=
\sum_{u=1}^{N_{\mathrm{user}}}
\lambda_k^{(u)} R_{k1}^{(u,n)},
\qquad 1 \le n \le N,
\label{eq:root_gamma}
\end{equation}
since a type-$k$ task generated by user $S_u$ sends its root subtask to server $S_n$ with probability $R_{k1}^{(u,n)}$.

Now consider a non-root subtask $v_{ki}$ whose parent is $v_{km}$. If the parent completes at an edge server $S_j \in \mathcal{E}$, then child $v_{ki}$ is sent to server $S_n$ with probability $R_{ki}^{(j,n)}$. Hence, for every child $v_{ki} \in C(v_{km})$,
\begin{equation}
\gamma_{ki,n}(R)
=
\sum_{S_j \in \mathcal{E}}
\gamma_{km,j}(R)\, R_{ki}^{(j,n)},
\qquad 1 \le n \le N.
\label{eq:child_gamma}
\end{equation}
Equation \eqref{eq:child_gamma} is the fundamental traffic equation of the model. It states that the batch-initiation rate of a child
at a destination server is obtained by summing its
parent-completing rates at the different edge servers, each
multiplied by the corresponding static routing probability.

Observe that descendants of a subtask sent to the cloud are already included in the aggregated quantity $D^{\mathrm{cloud}}_{km,n}$ (See \eqref{eq:DRkm}). Therefore, once a subtask is assigned to a cloud server, no additional downstream routing equations are needed for its descendants.

\subsection{Nominal Poisson Traffic at Each Server}

Applying the first step of the approximation to every initiated batch, the total nominal arrival rate of individual units to server $S_n$ is
\begin{equation}
\Lambda_n(R)
=
\sum_{k=1}^{K}
\sum_{m=1}^{M_k}
B_{km,n}\, \gamma_{km,n}(R).
\label{eq:Lambda_n}
\end{equation}
This is the total average rate of computation units or data units that reach server $S_n$ after replacing each batch by a Poisson stream with the same mean rate.

If one uses only \eqref{eq:Lambda_n}, then each node becomes an ordinary M/M/1 queue. However, this would ignore the fact that in the original system, the units within a batch arrive immediately one after another. Therefore, such a model would underestimate congestion.

\subsection{Congestion Correction}

To account for the additional congestion created by the fact that the
customers after the first one in each batch do not see the queue in steady
state, we add an independent Poisson stream of redundant customers at each
server. The role of this extra stream is only to reproduce the average congestion effect of the original batched traffic while keeping the queue quasi-reversible. Hence, when a redundant load completes service,
it leaves the system immediately and does not initiate any
subtask, does not generate any downstream batch, and does
not contribute to the traffic equations of subsequent queues.

The correction term is obtained by extending the single-queue argument in \cite{RahnamaniaAshtianiCorrelatedArrivals}. For a batch of size $B$, the first unit behaves as the reference Poisson arrival, while the remaining $B-1$ units arrive behind it and create additional waiting. Averaging the extra delay over all positions inside the batch produces the triangular contribution $B(B-1)/2$. Summing this over all batch types that feed server $S_n$, we define
\begin{equation}
C_n(R)
=
\sum_{k=1}^{K}
\sum_{m=1}^{M_k}
\frac{B_{km,n}\bigl(B_{km,n}-1\bigr)}{2}\,
\gamma_{km,n}(R).
\label{eq:C_n}
\end{equation}

Let $\alpha_n(R)$ denote the rate of the redundant Poisson stream added to server $S_n$. Its value is obtained by matching the mean number of customers in the original batch-arrived queue and in the approximating dual queue. The full derivation is given in Appendix~\ref{app:correction}. Here we only state the resulting relation:
\begin{equation}
\alpha_n(R)
=
\bigl(1-\rho_n(R)\bigr) C_n(R),
\qquad
\rho_n(R)
=
\frac{\Lambda_n(R)+\alpha_n(R)}{\mu_n}.
\label{eq:alpha_fixed}
\end{equation}
Solving \eqref{eq:alpha_fixed} gives
\begin{equation}
\rho_n(R)
=
\frac{\Lambda_n(R)+C_n(R)}{\mu_n+C_n(R)},
\label{eq:rho_closed}
\end{equation}
and
\begin{equation}
\alpha_n(R)
=
\frac{C_n(R)\bigl(\mu_n-\Lambda_n(R)\bigr)}{\mu_n+C_n(R)}.
\label{eq:alpha_closed}
\end{equation}

Therefore, each server is approximated by an M/M/1 queue with service rate $\mu_n$ and effective arrival rate $\Lambda_n(R)+\alpha_n(R)$.

\subsection{Waiting Time and Product-Form Approximation}

After the above transformation, all servers are modeled as quasi-reversible M/M/1 nodes driven by independent Poisson streams. Therefore, the whole system is approximated by a product-form queueing network. The average waiting time at server $S_n$ is then
\begin{equation}
W_n(R)
=
\frac{\rho_n(R)}{\mu_n\bigl(1-\rho_n(R)\bigr)}.
\label{eq:Wn}
\end{equation}

\subsection{Recursive Evaluation of Traffic and M/M/1 Parameters}

The traffic equations are solved in a topological order over the task tree.

\begin{enumerate}
    \item For each task type $k$ and each possible destination server $S_n$, compute the root initiation rates $\gamma_{k1,n}(R)$ from \eqref{eq:root_gamma}.
    \item Traverse the subtasks of $T_k$ in BFS order (or any topological order from parent to children). Whenever the rates $\gamma_{km,j}(R)$ of a parent $v_{km}$ are known, compute the rates of each child $v_{ki}\in C(v_{km})$ from \eqref{eq:child_gamma}.
    \item After all $\gamma_{km,n}(R)$ are obtained, compute for each server $S_n$ the nominal offered load $\Lambda_n(R)$ from \eqref{eq:Lambda_n} and the batch-correction term $C_n(R)$ from \eqref{eq:C_n}.
    \item Finally, obtain $\rho_n(R)$, $\alpha_n(R)$, and $W_n(R)$ from \eqref{eq:rho_closed}, \eqref{eq:alpha_closed}, and \eqref{eq:Wn}.
\end{enumerate}

Because the task graph is acyclic, no iterative fixed-point procedure is needed for the subtask traffic itself: the rates $\gamma_{km,n}(R)$ are obtained recursively from the root toward the leaves. Therefore, the full queueing-network approximation is computationally light and can be evaluated efficiently inside the optimization procedure.

It is important to interpret the proposed construction as a
local moment-matching approximation applied consistently across
the network. At each server, the original batched arrival
process is replaced by an equivalent Poisson flow with the same
mean rate, together with an additional Poisson as the flow whose
rate is chosen to recover the average congestion created by the
customers that arrive back-to-back within the same batch.
Since each node in the approximated regime is quasi-reversible, its
departure process is Poisson, which allows the same
transformation to be propagated to subsequent queues without
violating tractability. In this sense, the network model is
built by matching, queue by queue, the mean traffic intensity
and the mean congestion effect of batch correlation.

Accordingly, the approximation is not claimed to preserve the
full dependence structure of the original batch-arrived network
exactly. Its purpose is instead to capture the dominant delay
distortion induced by correlated internal batches while
retaining a product-form representation that can be evaluated
and optimized efficiently. For this reason, the approximation is
expected to be most reliable in the stable low-to-moderate load
regime, whereas some loss of accuracy is natural near
saturation, where higher-order dependence becomes more pronounced. The numerical
results in Section~\ref{subsec:numerical} confirm this behavior.


\section{Average Task Delay Derivation}
\label{sec:delay-derivation}

Our goal in this section is to express the average completion
delay of each task type in terms of the PFQN parameters 
derived earlier (initiation rates, utilizations and waiting times).
We first focus on a \emph{star-shaped} task whose graph has
one root and several children. Then we show how a general
task tree can be decomposed into stars and evaluated
recursively.

\begin{figure}[t]
  \centering
  \subfloat[]{
  \begin{tikzpicture}[
    >=Stealth,
    node distance=1.4cm and 1.6cm,
    task/.style={circle,draw,minimum size=6mm,inner sep=0pt},
    lab/.style={font=\scriptsize}
  ]

    \node[task] (k1) at (0,0) {$k1$};

    \node[task] (k2) at (2.0,-1.8) {$k2$};
    \node[task] (k3) at (2.0,-0.4) {$k3$};
    \node[task] (k4) at (2.0,1.2) {$k4$};

    \node[task] (k5) at (4.0,-0.4) {$k5$};

    \node[task] (k8) at (4.0,2.0) {$k8$};
    \node[task] (k7) at (4.0,1.2) {$k7$};
    \node[task] (k6) at (4.0,0.4) {$k6$};

    \draw[->] (k1) -- (k2);
    \draw[->] (k1) -- (k3);
    \draw[->] (k1) -- (k4);

    \draw[->] (k3) -- (k5);
    \draw[->] (k4) -- (k8);
    \draw[->] (k4) -- (k7);
    \draw[->] (k4) -- (k6);

    \draw[dashed] (1.4,0.05)  rectangle (4.6,2.4);  
    \draw[dashed] (1.4,-0.8) rectangle (4.6,-0.05);  

  \end{tikzpicture}
  }
  \hfill
  \subfloat[]{
  \begin{tikzpicture}[
    >=Stealth,
    node distance=1.4cm and 1.6cm,
    task/.style={circle,draw,minimum size=6mm,inner sep=0pt},
    lab/.style={font=\scriptsize}
  ]

    \node[task] (k1) at (0,0) {$k1$};

    \node[task] (k2) at (2.0,-1.4) {$k2$};
    \node[task] (k3) at (2.0,-0.4) {$k3$};
    \node[task] (k4) at (2.0,0.6) {$k4$};
    \node[task] (kD) at (2.0,2) {$kD$};

    \draw[->] (k1) -- (k2);
    \draw[->] (k1) -- (k3);
    \draw[->] (k1) -- (k4);
    \draw[->] (k1) -- (kD);

    \node[lab] at (2,1.3) {$\vdots$};

  \end{tikzpicture}
  }

  \caption{Star-based representation of task $T_k$ and its recursive
  decomposition: (a) decomposition of the task into stars and folding of
  their delays; (b) generic star used to define the delay parameters.}
  \label{fig:star-both}
\end{figure}
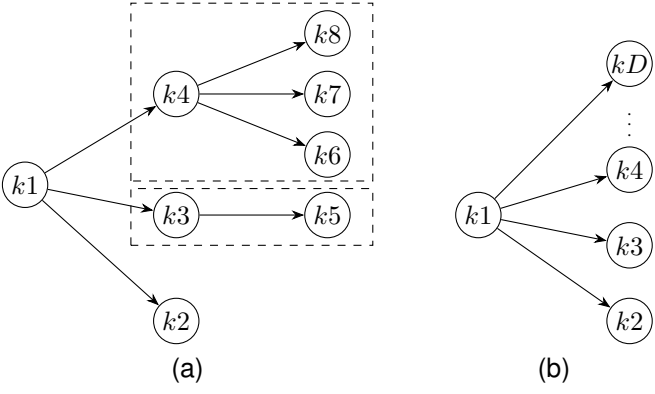

Fig.~\ref{fig:star-both} illustrates this construction. We show how a task is decomposed
into stars and collapsed from the leaves to the root, and a generic star used to define
the parameters in this section.

\subsection{Delay of a Star-Shaped Task}
\label{subsec:star-delay}

Consider a single star of task $T_k$ whose root is subtask
$v_{k1}$ and whose children are
$v_{k2},\ldots,v_{kD}$. We assume that the root $v_{k1}$
has just completed at server $S_n$, with $S_n \in \mathcal{E}$.
The children may be executed at different servers depending
on the static offloading policy $R$.
\newline
\textbf{Offloading scenarios.}
We enumerate all possible execution patterns for $v_{k2},\ldots,v_{kD}$ through the scenario set
\begin{equation}
  \mathcal{A}
  =
  \big\{
    a = (a_2,\ldots,a_D)
    \,\big|\,
    a_i \in \{1,\ldots,N\}
  \big\},
  \label{eq:scenario-set}
\end{equation}
where $a_i$ denotes the index of the server that executes
subtask $v_{ki}$. Under the static routing policy $R$, the
probability that scenario $a \in \mathcal{A}$ occurs given that
the parent $v_{k1}$ has been completed at $S_n$ is
\begin{equation}
  P_n(a)
  =
  \prod_{i=2}^{D}
  R_{ki}^{(n,a_i)},
  \label{eq:Pn-a}
\end{equation}
because the children are routed independently according to
the probabilities in \eqref{eq:child-normalization}.
\newline
\textbf{Edge and cloud processing times.}
For later use we group the average processing times of the
children into vectors. Define
\begin{equation}
  x
  =
  \big(
    x^{(k2)},\ldots,x^{(kD)}
  \big),
\end{equation}
where for each child $v_{ki}$ we set
\begin{equation}
  x^{(ki)}
  =
  \big(
    x^{(ki)}_1,\ldots,
    x^{(ki)}_{N_{\mathrm{user}}+N_{\mathrm{edge}}},
    x^{(ki)}_{N-N_{\mathrm{cloud}}+1},\ldots,
    x^{(ki)}_{N}
  \big),
  \label{eq:x-vector-def}
\end{equation}
with components
\begin{equation}
  x^{(ki)}_j
  =
  \begin{cases}
    \displaystyle
    \dfrac{l_{ki}}{\mu_j},
      & 1 \le j \le N_{\mathrm{user}} + N_{\mathrm{edge}},
      \\[1.2ex]
    \displaystyle
    \dfrac{D^{\mathrm{cloud}}_{ki,j}}{\mu_j'},
      & N - N_{\mathrm{cloud}} + 1 \le j \le N.
  \end{cases}
  \label{eq:x-components}
\end{equation}
The $j$-th element $x^{(ki)}_j$ represents the mean time
spent at server $S_j$ if $v_{ki}$
is processed there:
for edge servers it is the average computation time
$l_{ki}/\mu_j$, and for cloud servers it is the mean
transmission time of the aggregated data batch
$D^{\mathrm{cloud}}_{ki,j}$ defined in \eqref{eq:DRkm}.
\newline
\textbf{Contribution of a server in a given scenario.}
Given scenario $a=(a_2,\ldots,a_D)$, the average time that
the star spends at server $S_j$, $1 \le j \le N$, before all
children are completed can be written as
\begin{equation}
  \begin{aligned}
  T_j(a)
  &=
  W_j(R)\,
  \mathbf{1}\!\left(
    \exists\, i \in \{2,\ldots,D\} : a_i = j
  \right)
  \\
  &\quad+
  \sum_{i=2}^{D}
    x^{(ki)}_j\,
    \mathbf{1}(a_i = j),
  \end{aligned}
  \label{eq:Tj-a}
\end{equation}
where \(W_j(R)\) is the mean waiting time in the M/M/1 queue
at \(S_j\), and \(\mathbf{1}(\cdot)\) is the indicator function.
The first term accounts for the queueing delay and is included
only if at least one child is assigned to \(S_j\) in scenario
\(a\). When several children of the same star are assigned to
\(S_j\), they arrive as a single consecutive group and therefore
incur this queueing delay only once. The second term sums the
mean processing times of all children executed at \(S_j\) in
that scenario.
\newline
\textbf{Average star delay conditioned on the root server.}
The overall delay for completing the children
$v_{k2},\ldots,v_{kD}$ in scenario $a$ is governed by the
server that becomes the bottleneck. Since $T_j(a)$ is constructed
from mean queueing and processing times, we approximate the
conditional mean completion delay of the star by the maximum of
these mean server-wise delay contributions. Accordingly, when the
root $v_{k1}$ is completed at server $S_n$, the approximated star
delay is
\begin{equation}
  \mathrm{ET}_{k1}^{(n)}(R)
  =
  \begin{cases}
    \displaystyle
    \frac{l_{k1}}{\mu_n}
    +
    \sum_{a \in \mathcal{A}}
      \left[
        \max_{1 \le j \le N} T_j(a)
      \right]
      P_n(a),
      & S_n \in \mathcal{E},
      \\[2ex]
    \displaystyle
    \frac{D^{\mathrm{cloud}}_{k1,n}}{\mu_n'},
      & S_n \in \mathcal{C},
  \end{cases}
  \label{eq:ET-k1n}
\end{equation}
where $l_{k1}/\mu_n$ is the mean computation time of the
root when it is processed at an edge server $S_n$, and
$D^{\mathrm{cloud}}_{k1,n}/\mu_n'$ is the mean delay in the
cloud queue when the root and all its descendants are
offloaded to cloud server $S_n$.

For convenience we define the vector-valued mapping
\begin{equation}
  \mathrm{StarET}(k1,R,x)
  =
  \big(
    \mathrm{ET}_{k1}^{(1)}(R),
    \ldots,
    \mathrm{ET}_{k1}^{(N)}(R)
  \big),
  \label{eq:StarET-def}
\end{equation}
which returns the star delay for all possible servers of the
root, given the routing policy $R$ and the per-child vectors
$x$ in \eqref{eq:x-vector-def}.

\subsection{Recursive Delay Computation for a General Task}
\label{subsec:task-recursion}
We now extend the star analysis to a general task $T_k$ whose
computation graph is a rooted tree.
\begin{algorithm}[!t]
\caption{TaskET($T_k$, $R$): average delay of task $T_k$}
\label{alg:tasket}
\begin{algorithmic}[1]
\STATE \textbf{function} SubtaskET$_k(m)$
\IF{$C(v_{km}) = \emptyset$}
    \STATE \{ $v_{km}$ is a leaf \}
    \STATE construct $x^{(km)}$ as in \eqref{eq:x-vector-def}
    \STATE \textbf{return} $x^{(km)}$
\ELSE
    \FOR{each child $v_{ki} \in C(v_{km})$}
        \STATE $x^{(ki)} \gets$ SubtaskET$_k(i)$
    \ENDFOR
    \STATE $x \gets \left(x^{(ki)} : v_{ki} \in C(v_{km})\right)$
    \STATE \textbf{return} StarET$(km, R, x)$
\ENDIF
\STATE \textbf{end function}
\STATE
\STATE $(ET_{k1}^{(1)}, \ldots, ET_{k1}^{(N)}) \gets$ SubtaskET$_k(1)$
\STATE TaskET$(T_k, R) \gets
\dfrac{
\sum_{n=1}^{N}
\gamma_{k1,n}(R)\left(W_n(R)+ET_{k1}^{(n)}(R)\right)
}{
\sum_{u=1}^{N_{\mathrm{user}}}\lambda_k^{(u)}
}$
\end{algorithmic}
\end{algorithm}
Algorithm~\ref{alg:tasket} evaluates the task tree from the
leaves toward the root. For a leaf subtask, it returns the vector
of its mean execution delays over all possible servers. For an
internal subtask, it first evaluates all of its children and then
uses \(\mathrm{StarET}\) to collapse the corresponding star into
a single server-dependent delay vector. Repeating this procedure
recursively yields the delay vector of the root subtask. Finally,
the algorithm averages the root delays over its possible execution
servers using the corresponding batch-initiation rates and adds
the waiting time at the root server.
\subsection{Network-Wide Average Delay}

Finally, we obtain the average delay across all task types by
weighting the per-type delays $\mathrm{TaskET}(T_k,R)$ with
their aggregate arrival rates. Let
$\lambda_k = \sum_{n=1}^{N_{\mathrm{user}}} \lambda_k^{(n)}$ be
the total arrival rate of type-$k$ tasks. The global metric is
\begin{equation}
  T_{\mathrm{tot}}(R)
  =
  \frac{
    \displaystyle
    \sum_{k=1}^{K}
      \lambda_k\,
      \mathrm{TaskET}(T_k,R)
  }{
    \displaystyle
    \sum_{k=1}^{K} \lambda_k
  }.
  \label{eq:Ttot-final}
\end{equation}
This is the objective function that is minimized in the next
section by appropriately selecting the static routing
probabilities in $R$.

\section{Optimization of the Static Offloading Policy}
\label{sec:optimization}

The routing decision for a root subtask is encoded by
\(R_{k1}^{(n_0,n)}\), where \(S_{n_0}\) is the user that
generates the task. For a non-root subtask, the routing decision
is encoded by \(R_{km}^{(n_p,n)}\), where \(S_{n_p}\) is the
server at which its parent subtask completes.

\subsection{Problem Formulation}
Using the delay metric derived in Section~\ref{sec:delay-derivation}, we seek the
static routing policy that minimizes the network-wide average
task delay:
\begin{equation}
\begin{aligned}
\min_{R} \quad
  & T_{\mathrm{tot}}(R) \\
\text{s.t.} \quad
  & \rho_n(R)<1,
  && 1\le n\le N,\\
  & \sum_{S_n\in\mathcal{Z}_{n_0}}
    R_{k1}^{(n_0,n)}=1,
  && \forall\,k,\ 1\le n_0\le N_{\mathrm{user}},\\
  & R_{k1}^{(n_0,n)}\ge 0,
  && \forall\,k,n_0,\ S_n\in\mathcal{Z}_{n_0},\\
  & R_{k1}^{(n_0,n)}=0,
  && \forall\,k,n_0,\ S_n\notin\mathcal{Z}_{n_0},\\
  & \sum_{S_n\in\mathcal{Z}_{n_p}}
    R_{km}^{(n_p,n)}=1,
  && \forall\,k,\ 2\le m\le M_k,\
     S_{n_p}\in\mathcal{E},\\
  & R_{km}^{(n_p,n)}\ge 0,
  && \forall\,k,\ 2\le m\le M_k,\
   S_{n_p}\in\mathcal{E},\\
  & && S_n\in\mathcal{Z}_{n_p},\\
  & R_{km}^{(n_p,n)}=0,
  && \forall\,k,\ 2\le m\le M_k,\
   S_{n_p}\in\mathcal{E},\\
  & && S_n\notin\mathcal{Z}_{n_p}.
\end{aligned}
\label{eq:opt-problem}
\end{equation}
Here $\rho_n(R)$ is the utilization of approximated queue corresponding to server
$S_n$, obtained from the queueing approximation in Section~\ref{sec:qn_model}:
\[
  \rho_n(R)
  =
  \frac{\Lambda_n(R) + \alpha_n(R)}{\mu_n},
\]
where $\Lambda_n(R)$ is the nominal Poisson traffic at server
$S_n$, and $\alpha_n(R)$ is the additional Poisson correction
term that captures the congestion effect of back-to-back
arrivals within the batches.

\subsection{Softmax Parameterization of Routing Probabilities}
\label{subsec:softmax}

Instead of enforcing the probability constraints in
\eqref{eq:opt-problem} directly, we introduce unconstrained
\emph{logit} variables $\theta$ and express the routing probabilities
through softmax mappings.

For each root-routing tuple \((k,n_0)\), we define
\begin{equation}
R_{k1}^{(n_0,n)}(\theta)
=
\begin{cases}
\displaystyle
\frac{
\exp\!\bigl(\theta_{k1}^{(n_0,n)}\bigr)
}{
\sum_{S_j\in\mathcal{Z}_{n_0}}
\exp\!\bigl(\theta_{k1}^{(n_0,j)}\bigr)
},
& S_n\in\mathcal{Z}_{n_0},\\[2ex]
0,
& S_n\notin\mathcal{Z}_{n_0}.
\end{cases}
\label{eq:softmax-root}
\end{equation}

For each non-root routing tuple \((k,m,S_{n_p})\), where
\(2\le m\le M_k\) and \(S_{n_p}\in\mathcal{E}\), we define
\begin{equation}
R_{km}^{(n_p,n)}(\theta)
=
\begin{cases}
\displaystyle
\frac{
\exp\!\bigl(\theta_{km}^{(n_p,n)}\bigr)
}{
\sum_{S_j\in\mathcal{Z}_{n_p}}
\exp\!\bigl(\theta_{km}^{(n_p,j)}\bigr)
},
& S_n\in\mathcal{Z}_{n_p},\\[2ex]
0,
& S_n\notin\mathcal{Z}_{n_p}.
\end{cases}
\label{eq:softmax-child}
\end{equation}

By construction, both the root-routing and child-routing
probability vectors lie on their corresponding probability
simplices, so all routing constraints are automatically
satisfied. All queueing parameters derived in Section III, namely
$\gamma_{km,n}(\theta)$, $\Lambda_n(\theta)$, $C_n(\theta)$,
$\alpha_n(\theta)$, $\rho_n(\theta)$, and $W_n(\theta)$, become
differentiable functions of $\theta$.

We denote the resulting objective by
\[
  T_{\mathrm{tot}}(\theta)
  =
  T_{\mathrm{tot}}\big(R(\theta)\big),
\]
and seek to minimize it over $\theta$.

\subsection{Smooth Approximation of Star Maxima}
\label{subsec:smooth-max}

The star delay expression \eqref{eq:ET-k1n} contains the
pointwise maximum $\max_j T_j(a)$, which makes
$T_{\mathrm{tot}}(\theta)$ non-smooth. To obtain a
differentiable surrogate we replace this maximum by a
log-sum-exp (LSE) smoothing~\cite{BoydConvex}. For a fixed scenario $a$ and
parameter $\tau>0$ we define
\begin{equation}
  \mathrm{LSE}_\tau\big(T_1(a),\ldots,T_N(a)\big)
  =
  \frac{1}{\tau}
  \log \left(
    \sum_{j=1}^{N}
      \exp\!\big( \tau\, T_j(a) \big)
  \right),
  \label{eq:LSE-def}
\end{equation}
which satisfies
\[
  \max_j T_j(a)
  \le
  \mathrm{LSE}_\tau(\cdot)
  \le
  \max_j T_j(a) + \frac{\log N}{\tau}.
\]
Replacing $\max_j T_j(a)$ by $\mathrm{LSE}_\tau$ in
\eqref{eq:ET-k1n} yields a smooth approximation
$\widetilde{\mathrm{ET}}_{k1}^{(n)}(\theta)$ and, through
Algorithm~\ref{alg:tasket}, smooth approximations
$\widetilde{\mathrm{TaskET}}(T_k,\theta)$ and
$\widetilde{T}_{\mathrm{tot}}(\theta)$ to the original delay
metrics. For an appropriate $\tau$ this approximation is
accurate while keeping the gradients numerically stable,
in line with standard smooth-max techniques in convex
optimization~\cite{BoydConvex,BertsekasNonlinear}.

\subsection{Stability Barrier on Server Utilizations}
\label{subsec:rho-barrier}

To enforce the stability constraints $\rho_n(R) < 1$ after the
softmax reparameterization, we add a logarithmic barrier on
the server utilizations computed by the queueing model. Let
$\rho_n(\theta)$ be the utilization of server
$S_n$ under routing $R(\theta)$, and fix a safety level
$\rho_{\max}<1$ (e.g., $\rho_{\max}=0.99$). We define
\begin{equation}
B(\theta)=\kappa\sum_{n=1}^{N} b\bigl(\rho_n(\theta)\bigr),
\qquad
b(\rho)=-\log\!\left(1-\frac{\rho}{\rho_{\max}}\right),
\label{eq:barrier}
\end{equation}
with weight $\kappa>0$. The term $B(\theta)$ grows to $+\infty$
as any $\rho_n(\theta)$ approaches $\rho_{\max}$, so gradient
iterations remain in the interior of the stability region. This construction follows standard barrier and interior-point methods
for constrained optimization~\cite{BoydConvex,BertsekasNonlinear}.

The objective used in the numerical optimization is therefore
\begin{equation}
\widetilde{T}_{\mathrm{tot}}^{\,b}(\theta)
=
\widetilde{T}_{\mathrm{tot}}(\theta)+B(\theta),
\label{eq:barrier_obj}
\end{equation}
which is smooth in $\theta$ as long as
$\rho_n(\theta) < \rho_{\max}$ for all $n$.

\subsection{Gradient-Based Optimization Algorithm}
\label{subsec:gradient-algorithm}

Given the smooth, barrier-augmented objective
$\widetilde{T}_{\mathrm{tot}}^{\,b}(\theta)$, we apply a standard
gradient-descent scheme~\cite{BertsekasNonlinear}
to obtain a locally optimal static offloading policy, as
summarized in Algorithm~\ref{alg:gradient-opt}.

\begin{algorithm}
\caption{Gradient-based optimization of static offloading}
\label{alg:gradient-opt}
\begin{algorithmic}[1]
\STATE choose initial logits $\theta^{(0)}$
\STATE choose smoothing parameter $\tau > 0$, step-size rule
$\{\eta^{(t)}\}_{t \ge 0}$, and tolerance $\varepsilon > 0$
\STATE $t \gets 0$
\REPEAT
\STATE build routing matrix $R(\theta^{(t)})$ using
\eqref{eq:softmax-root} and \eqref{eq:softmax-child}
\STATE solve the recursive traffic equations to obtain
$\gamma_{km,n}(\theta^{(t)})$
\STATE compute $\Lambda_n(\theta^{(t)})$ from \eqref{eq:Lambda_n}
and $C_n(\theta^{(t)})$ from \eqref{eq:C_n}
\STATE compute $\alpha_n(\theta^{(t)})$ from \eqref{eq:alpha_closed},
$\rho_n(\theta^{(t)})$ from \eqref{eq:rho_closed}, and
$W_n(\theta^{(t)})$ from \eqref{eq:Wn}
\STATE evaluate smoothed star delays
$\widetilde{ET}_{km}^{(n)}(\theta^{(t)})$ using \eqref{eq:Tj-a} and \eqref{eq:LSE-def}
\STATE run Algorithm~\ref{alg:tasket} with the smoothed star delays to obtain
$\widetilde{T}_{\mathrm{tot}}(\theta^{(t)})$
\STATE compute the barrier term $B(\theta^{(t)})$
using \eqref{eq:barrier}
\STATE form total objective
$\widetilde{T}_{\mathrm{tot}}^{\,b}(\theta^{(t)})
\gets
\widetilde{T}_{\mathrm{tot}}(\theta^{(t)})+B(\theta^{(t)})$
\STATE compute gradient
$G^{(t)} \gets \nabla_{\theta}
\widetilde{T}_{\mathrm{tot}}^{\,b}(\theta^{(t)})$
\STATE update logits
$\theta^{(t+1)} \gets \theta^{(t)}-\eta^{(t)}G^{(t)}$
\STATE $t \gets t+1$
\UNTIL{$\left|
\widetilde{T}_{\mathrm{tot}}^{\,b}(\theta^{(t)})
-
\widetilde{T}_{\mathrm{tot}}^{\,b}(\theta^{(t-1)})
\right| \le \varepsilon$}
\STATE $R^\star \gets R(\theta^{(t)})$
\end{algorithmic}
\end{algorithm}

\section{Performance Evaluation}
\label{subsec:numerical}

In this section, we evaluate the proposed framework from three complementary perspectives. First, we assess the accuracy of the queueing-network approximation. Second, we compare the delay performance of the optimized policy with the baseline algorithms. Third, we evaluate their maximum stable throughput under mixed and single-class traffic.

\subsection{System Configuration and Evaluation Methodology}

We emulate a canonical MEC deployment in which a set of
resource-constrained user devices share a powerful edge server and
a remote cloud queue. Concretely, we consider $N=4$ single-server
FCFS queues, indexed by
\[
S=\{S_1,S_2,S_3,S_4\},
\]
and interpreted as two user devices ($S_1,S_2$), one edge/MEC
server ($S_3$), and one cloud server ($S_4$). The service
rates are
\begin{equation}
\mu=(\mu_1,\mu_2,\mu_3,\mu_4')=(0.6,0.7,3.7,2.2).
\label{eq:eval_mu_final}
\end{equation}

The connection zones are
\[
  Z_1 = \{1,2,3,4\},\hspace{0.1cm}
  Z_2 = \{1,2,3\},\hspace{0.1cm}
  Z_3 = \{1,2,3,4\},\hspace{0.1cm}
  Z_4 = \{4\}.
\]
The purpose of this configuration is not to emulate a large-scale
deployment, but to provide a limited setting that contains
all essential tiers of the considered MEC architecture: mobile user
devices, a shared edge server, and a remote cloud queue. The choice
$N=4$ is the smallest nontrivial topology that simultaneously captures
local execution, edge offloading, cloud offloading, heterogeneous service
rates, and constrained connectivity. This compact setting also makes it
possible to isolate the effect of the proposed batch-corrected queueing
approximation and to compare analytical and simulated delays over many
routing policies and offered-load values without confounding the results
with large-network topology effects.

\begin{figure*}[t]
  \centering
  \begin{tikzpicture}[
    >=Stealth,
    node distance=1.4cm and 1.6cm,
    task/.style={circle,draw,minimum size=5.5mm,inner sep=0pt},
    lab/.style={font=\scriptsize},
    edgelab/.style={font=\scriptsize,inner sep=1pt}
  ]

  \begin{scope}[xshift=0cm]
    \node[task] (t11) at (0,0) {$11$};
    \node[lab,above left=0mm and -0.5mm of t11]
      {$\ell_{11}=2$};
    \node[lab,below left=0mm and -0.5mm of t11]
      {$d_{(\mathrm{out},11)}=1$};

    \node[task] (t13) at (2.0,0.9) {$13$};
    \node[lab,above right=0mm and -0.5mm of t13]
      {$\ell_{13}=2$};
    \node[lab,below right=0mm and -0.5mm of t13]
      {$r_{13}=1$};

    \node[task] (t12) at (2.0,-0.9) {$12$};
    \node[lab,above right=0mm and -0.5mm of t12]
      {$\ell_{12}=3$};
    \node[lab,below right=0mm and -0.5mm of t12]
      {$r_{12}=1$};

    \draw[->] (t11) --
      node[edgelab,sloped,above] {$d_{(11,13)}=2$} (t13);
    \draw[->] (t11) --
      node[edgelab,sloped,below] {$d_{(11,12)}=1$} (t12);
  \end{scope}

  \begin{scope}[xshift=5.7cm]
    \node[task] (t21) at (0,0.2) {$21$};
    \node[lab,above left=0mm and -0.5mm of t21]
      {$\ell_{21}=2$};
    \node[lab,below left=0mm and -0.5mm of t21]
      {$d_{(\mathrm{out},21)}=1$};

    \node[task] (t23) at (2.0,1.1) {$23$};
    \node[lab,above right=0mm and -0.5mm of t23]
      {$\ell_{23}=2$};
    \node[lab,below right=0mm and -0.5mm of t23]
      {$r_{23}=1$};

    \node[task] (t22) at (2.0,-0.8) {$22$};
    \node[lab,below=1mm of t22] {$\ell_{22}=3$};

    \node[task] (t25) at (4.0,0.2) {$25$};
    \node[lab,above=0mm of t25] {$\ell_{25}=2$};
    \node[lab,below=0mm of t25] {$r_{25}=1$};

    \node[task] (t24) at (4.0,-1.6) {$24$};
    \node[lab,above=0mm of t24] {$\ell_{24}=4$};
    \node[lab,below=0mm of t24] {$r_{24}=1$};

    \draw[->] (t21) --
      node[edgelab,sloped,above] {$d_{(21,23)}=3$} (t23);
    \draw[->] (t21) --
      node[edgelab,sloped,below] {$d_{(21,22)}=1$} (t22);
    \draw[->] (t22) --
      node[edgelab,sloped,above] {$d_{(22,25)}=2$} (t25);
    \draw[->] (t22) --
      node[edgelab,sloped,below] {$d_{(22,24)}=1$} (t24);
  \end{scope}

  \begin{scope}[xshift=11.4cm]
    \node[task] (t31) at (0,1) {$31$};
    \node[lab,above=1mm of t31] {$\ell_{31}=1$};
    \node[lab,below=0.5mm of t31]
      {$d_{(\mathrm{out},31)}=2$};

    \node[task] (t32) at (2.5,1.0) {$32$};
    \node[lab,above right=0mm and -0.5mm of t32]
      {$\ell_{32}=2$};

    \node[task] (t33) at (0,-1.5) {$33$};
    \node[lab,below=0mm of t33] {$\ell_{33}=3$};

    \node[task] (t34) at (2.5,-1.5) {$34$};
    \node[lab,above right=0mm and -0.5mm of t34]
      {$\ell_{34}=4$};
    \node[lab,below right=0mm and -0.5mm of t34]
      {$r_{34}=1$};

    \draw[->] (t31) --
      node[edgelab,sloped,above] {$d_{(31,32)}=1$} (t32);
    \draw[->] (t32) --
      node[edgelab,sloped,below] {$d_{(32,33)}=1$} (t33);
    \draw[->] (t33) --
      node[edgelab,sloped,above] {$d_{(33,34)}=1$} (t34);
  \end{scope}

  \end{tikzpicture}
  \caption{Task graphs used in the numerical experiments, with
  computation loads $\ell_m$, result loads $r_m$, and inter-subtask
  data sizes $d_{(m,m')}$ for each edge.}
  \label{fig:three-tasks}
\end{figure*}
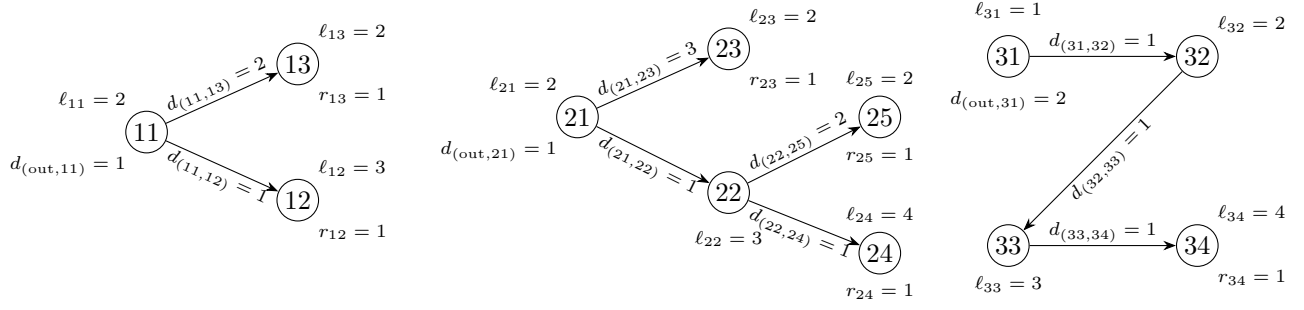

We consider three task types $T_1,T_2,T_3$, each modeled as a
rooted directed tree.
The three trees are chosen to capture a shallow fork--join
pattern, a hub--and--spoke pattern, and a deeper skewed tree
system~\cite{TaskGraphMEC}. All loads $(\ell_m,r_m)$ and data sizes
$d_{(m,m')}$ are shown in Fig.~\ref{fig:three-tasks}.

Let $\Lambda_{\mathrm{sum}}$ denote the total external arrival rate of
tasks. For the mixed-traffic delay experiment, it is swept over
\begin{equation}
  \Lambda_{\mathrm{sum}}
  \in
  \{0.10,0.20,0.30,\ldots,1.20\}\,\text{tasks/s}.
\label{eq:sweep}
\end{equation}
For each value of $\Lambda_{\mathrm{sum}}$, the three task types are
generated according to independent Poisson processes with equal rates
\begin{equation}
  \lambda_k
  =
  \frac{\Lambda_{\mathrm{sum}}}{3},
  \qquad k \in \{1,2,3\}.
\end{equation}
The aggregate type-$k$ rate is split equally
between the two mobile users, i.e.,
\[
    \lambda_k^{(1)}=\lambda_k^{(2)}=\frac{\lambda_k}{2}.
\]
The offered-load sweep in \eqref{eq:sweep} is designed to evaluate the
model over a wide range of queueing regimes. As reported later in
Table~\ref{tab:server-utilization}, the resulting simulated utilizations
$\rho_n$ span light-load, moderate-load, and near-saturation operating
points across different servers and total arrival rates. Therefore, the
experiments do not test the proposed approximation only at a single
lightly loaded point; they expose it to a broad set of utilization levels
and bottleneck patterns.

For each experiment, the analytical parameters are obtained from
the PFQN approximation developed in
Section~\ref{sec:delay-derivation}, and the corresponding
ground-truth delays and utilizations are obtained by discrete-event
simulation.

\subsection{Accuracy of the Queueing Approximation}

We first assess the accuracy of the proposed queueing-network
approximation. To this
end, we generate $20$ feasible static routing policies by
randomizing the routing probabilities over the admissible connection
zones. We then retain only those policies whose simulated maximum
server utilization satisfies
\[
\max_n \rho_n^{\mathrm{sim}} < 0.98.
\]
This yields a stable random-policy ensemble over which the analytical
delay predictions can be compared with discrete-event simulation.

Under this stability criterion, all $20$ sampled policies are retained
for
\[
\Lambda_{\mathrm{sum}}
\in
\{0.05,0.10,0.15,0.20\}.
\]
Thus, the approximation-accuracy study below uses the same number of
stable policies at every tested load level.

Table~\ref{tab:approximation_accuracy} reports the resulting delay-prediction
errors. In Table~\ref{tab:approximation_accuracy}, ``corr.'' denotes the
approximation with congestion correction, while ``uncorr.'' denotes
the approximation without congestion correction.
The approximation with congestion correction remains reasonably
accurate over the stable random-policy set. Its mean relative error is
$2.85\%$ at $\Lambda_{\mathrm{sum}}=0.05$,
$4.57\%$ at $\Lambda_{\mathrm{sum}}=0.10$,
$7.06\%$ at $\Lambda_{\mathrm{sum}}=0.15$, and
$8.37\%$ at $\Lambda_{\mathrm{sum}}=0.20$. In contrast, the
approximation without congestion correction is substantially less
accurate, with mean relative errors of $8.20\%$, $15.99\%$,
$26.12\%$, and $35.55\%$ at the same load levels. Hence, the
congestion-correction step provides a clear and systematic
improvement over the pure Poissonized approximation.

\begin{table*}[!t]
\caption{Approximation accuracy over stable random routing policies.}
\label{tab:approximation_accuracy}
\centering
\renewcommand{\arraystretch}{1.1}
\setlength{\tabcolsep}{6pt}
\begin{tabular}{c c c c c c}
\hline
$\Lambda_{\mathrm{sum}}$
& \begin{tabular}[c]{@{}c@{}}Mean delay\\(sim.)\end{tabular}
& \begin{tabular}[c]{@{}c@{}}Mean delay\\(corr.)\end{tabular}
& \begin{tabular}[c]{@{}c@{}}Mean delay\\(uncorr.)\end{tabular}
& \begin{tabular}[c]{@{}c@{}}Mean rel. err.\\(corr.)\end{tabular}
& \begin{tabular}[c]{@{}c@{}}Var. rel. err.\\(corr.) $(\times 10^{-4})$\end{tabular}
\\
\hline
0.05 & 8.6432  & 8.3944  & 7.9284  & 0.0285 & 0.449 \\
0.10 & 10.5975 & 10.1086 & 8.8857  & 0.0457 & 0.846 \\
0.15 & 14.8552 & 13.7782 & 10.8679 & 0.0706 & 2.822 \\
0.20 & 25.9426 & 23.8703 & 16.2526 & 0.0837 & 5.244 \\
\hline
\end{tabular}
\end{table*}

To put these errors in context, Table~\ref{tab:randutil_new}
reports the average simulated utilizations over the same stable
random-policy set. As expected, the approximation error increases as
the system moves closer to saturation. 

\begin{table}[t]
\caption{Average simulated utilizations over stable random routing policies.}
\label{tab:randutil_new}
\centering
\begin{tabular}{c c c c c}
\hline
$\Lambda_{\mathrm{sum}}$ & $\rho_1$ & $\rho_2$ & $\rho_3$ & $\rho_4$ \\
\hline
0.05 & 0.1853 & 0.1664 & 0.0301 & 0.0331 \\
0.10 & 0.3814 & 0.3251 & 0.0631 & 0.0648 \\
0.15 & 0.5922 & 0.4812 & 0.0920 & 0.0972 \\
0.20 & 0.7493 & 0.6863 & 0.1265 & 0.1290 \\
\hline
\end{tabular}
\end{table}

Overall, these results support two conclusions. First, the proposed
queueing-network approximation remains accurate over a broad set of
stable random routing policies in the low-to-moderate load regime.
Second, the congestion-correction step is essential: without it, the
analytical model substantially underestimates the delay induced by
the within-batch dependence.

\subsection{Performance of the Optimized Static Policy}

Having validated the analytical approximation over a broad family of
stable routing policies, we now evaluate the optimized static
offloading policy obtained from Section~\ref{sec:optimization}. We
compare it against four baselines:

\begin{itemize}

\item \textbf{URO (Uniform Random Offloading):}
For a non-root subtask, each feasible destination in the
connection zone of the parent-completion server is chosen with
equal probability. Thus,
\begin{equation}
R_{km}^{\mathrm{URO},(n_p,n)}
=
\begin{cases}
\dfrac{1}{|\mathcal{Z}_{n_p}|},
& S_n\in\mathcal{Z}_{n_p},\\[2mm]
0,
& S_n\notin\mathcal{Z}_{n_p},
\end{cases}
\qquad 2\le m\le M_k.
\label{eq:uro_policy}
\end{equation}
For a root subtask generated at user $S_{n_0}$, the same rule
is applied using the connection zone $\mathcal{Z}_{n_0}$.
Hence, URO ignores both server speed and congestion, and only respects
the connectivity constraint.

\item \textbf{PO (Proportional-to-capacity Offloading):}
For a non-root subtask ($2\le m\le M_k$), each feasible destination is selected
with probability proportional to its service rate. Thus,
\begin{equation}
R_{km}^{\mathrm{PO},(n_p,n)}
=
\begin{cases}
\displaystyle
\dfrac{\mu_n}
{\sum_{S_j\in\mathcal{Z}_{n_p}}\mu_j},
& S_n\in\mathcal{Z}_{n_p},\\[3mm]
0,
& S_n\notin\mathcal{Z}_{n_p},
\end{cases}.
\label{eq:po_policy}
\end{equation}
For a root subtask generated at user $S_{n_0}$, the same rule
is applied using the connection zone $\mathcal{Z}_{n_0}$.
Therefore, PO is still static, but biases the routing decisions
toward faster servers.

\item \textbf{ITAGS (Incremental Task-Aware Greedy Scheduling):}
This baseline is implemented as a dynamic greedy
scheduler~\cite{SundarLiang2018}. When a child subtask $v_{ki}$
becomes ready at time $t$ after the completion of its parent at
server $S_{n_p}$, every feasible destination
$S_n\in\mathcal{Z}_{n_p}$ is evaluated through the estimated
completion-time metric
\begin{equation}
\widehat{C}_{ki}^{\mathrm{ITAGS}}(n,t)
=
\widehat{W}_n(t)+b_{ki,n},
\label{eq:itags_metric}
\end{equation}
where $\widehat{W}_n(t)$ is the current queueing-delay estimate at
server $S_n$ derived from its instantaneous queue length in the
simulator, and
\begin{equation}
b_{ki,n}=
\begin{cases}
\dfrac{l_{ki}}{\mu_n}, & S_n \in \mathcal{E},\\[2mm]
\dfrac{D_{ki,n}^{\mathrm{cloud}}}{\mu_n'}, & S_n \in \mathcal{C}.
\end{cases}
\label{eq:itags_service_term}
\end{equation}
The subtask is then routed greedily to
\begin{equation}
n_{ki}^{\star}(t)
=
\arg\min_{S_n\in\mathcal{Z}_{n_p}}
\widehat{C}_{ki}^{\mathrm{ITAGS}}(n,t),
\label{eq:itags_argmin}
\end{equation}
that is,
\begin{equation}
R_{ki}^{\mathrm{ITAGS},(n_p,n)}(t)
=
\mathbf{1}\!\left(n=n_{ki}^{\star}(t)\right).
\label{eq:itags_policy}
\end{equation}
Thus, ITAGS makes a new routing decision whenever a subtask becomes
ready, using the instantaneous queue state.

\item \textbf{EFO (Expected Flow-Optimal Offloading):}
This baseline is implemented as a dynamic one-shot task-level
offloading policy~\cite{ShuZhaoHanMinDuan2019}. When a type-$k$
task arrives at user $S_{n_0}$ at time $t$, each feasible destination
$S_n \in \mathcal{Z}_{n_0}$ is evaluated through the task-level metric
\begin{equation}
\widehat{C}_{k}^{\mathrm{EFO}}(n,t)
=
\widehat{W}_n(t)+B_{k,n}^{\mathrm{tot}},
\label{eq:efo_metric}
\end{equation}
where $\widehat{W}_n(t)$ is the current queueing-delay estimate and
$B_{k,n}^{\mathrm{tot}}$ is the total workload of task $T_k$ if the
whole task is assigned to server $S_n$. The selected destination is
\begin{equation}
n_k^{\star}(t)
=
\arg\min_{n \in \mathcal{Z}_{n_0}}
\widehat{C}_{k}^{\mathrm{EFO}}(n,t),
\label{eq:efo_argmin}
\end{equation}
and all subtasks of the task are then executed at that same server:
\begin{equation}
R_{km}^{\mathrm{EFO},(n_0,n)}(t)
=
\mathbf{1}\!\left(n=n_k^{\star}(t)\right),
\quad 1 \le m\le M_k.
\label{eq:efo_policy}
\end{equation}
Hence, EFO is dynamic at the task level, but does not revise the
decision as the task progresses through its dependency tree.

\end{itemize}

Having defined the baseline policies, we now compare them with the
PFQN-based optimized static policy.
In the optimization step, we use a softmax parameterization with
smoothing parameter $\tau=10.0$, learning rate $0.05$, $250$
optimization epochs, utilization cap $\rho_{\max}=0.999$, penalty
weight $500.0$, and entropy regularization is included to discourage
prematurely concentrated routing distributions and improve the
numerical behavior of softmax-based policy optimization
~\cite{Ahmed2019Entropy,Mei2020Softmax}. For each
arrival-rate point, the PFQN-based policy and each baseline are
evaluated by $25$ independent simulation runs.

\begin{figure}[t]
\centering
\begin{tikzpicture}
\begin{axis}[
    width=\columnwidth,
    height=0.80\columnwidth,
    xmin=0.1,
    xmax=1.2,
    xtick={0.1,0.3,0.5,0.7,0.9,1.1,1.2},
    xlabel={Total arrival rate $\Lambda_{\mathrm{sum}}$},
    ylabel={Rate-weighted average task delay (s)},
    ymode=log,
    log basis y=10,
    ymin=1,
    ymax=5000,
    grid=both,
    major grid style={dashed,gray!35},
    minor grid style={dotted,gray!20},
    tick label style={font=\scriptsize},
    label style={font=\small},
    mark size=2.1pt,
    line width=0.9pt,
    legend style={
        at={(0.5,-0.28)},
        anchor=north,
        legend columns=2,
        draw=none,
        fill=none,
        font=\scriptsize,
        /tikz/every even column/.append style={column sep=0.55em}
    }
]

\addplot[
    color=PFQNBlue,
    thick,
    solid,
    mark=triangle*,
    mark options={draw=PFQNBlue,fill=PFQNBlue}
]
coordinates {
    (0.1,1.657196)
    (0.2,1.813371)
    (0.3,2.029210)
    (0.4,2.262831)
    (0.5,2.562129)
    (0.6,2.996822)
    (0.7,3.658168)
    (0.8,4.488958)
    (0.9,5.979422)
    (1.0,7.569675)
    (1.1,11.053894)
    (1.2,19.752408)
};
\addlegendentry{PFQN-based (simulation)}

\addplot[
    color=PFQNCyan,
    thick,
    dashed,
    mark=triangle,
    mark options={draw=PFQNCyan,fill=white}
]
coordinates {
    (0.1,1.629859)
    (0.2,1.784169)
    (0.3,1.983780)
    (0.4,2.195985)
    (0.5,2.460594)
    (0.6,2.813144)
    (0.7,3.314736)
    (0.8,4.091837)
    (0.9,5.463378)
    (1.0,6.732954)
    (1.1,9.533485)
    (1.2,17.607324)
};
\addlegendentry{PFQN-based (analytical)}

\addplot[
    color=ITAGSGreen,
    thick,
    solid,
    mark=diamond*,
    mark options={draw=ITAGSGreen,fill=ITAGSGreen}
]
coordinates {
    (0.1,2.513386)
    (0.2,2.684611)
    (0.3,2.909773)
    (0.4,3.175881)
    (0.5,3.542328)
    (0.6,3.998063)
    (0.7,4.687142)
    (0.8,5.788084)
    (0.9,7.992730)
    (1.0,17.782865)
    (1.1,163.285620)
    (1.2,376.682291)
};
\addlegendentry{ITAGS}

\addplot[
    color=EFOPurple,
    thick,
    dotted,
    mark=pentagon*,
    mark options={draw=EFOPurple,fill=EFOPurple}
]
coordinates {
    (0.1,2.286165)
    (0.2,2.544212)
    (0.3,2.904658)
    (0.4,3.478409)
    (0.5,4.475459)
    (0.6,6.412296)
    (0.7,9.944599)
    (0.8,14.324391)
    (0.9,20.186639)
    (1.0,42.411243)
    (1.1,170.948474)
    (1.2,248.392231)
};
\addlegendentry{EFO}

\addplot[
    color=PORed,
    thick,
    dashdotted,
    mark=square*,
    mark options={draw=PORed,fill=PORed}
]
coordinates {
    (0.1,4.326138)
    (0.2,4.795082)
    (0.3,5.479346)
    (0.4,6.426912)
    (0.5,7.791230)
    (0.6,10.412217)
    (0.7,15.190224)
    (0.8,32.458928)
    (0.9,162.141295)
    (1.0,410.850773)
    (1.1,592.665692)
    (1.2,734.136382)
};
\addlegendentry{PO}

\addplot[
    color=UROOrange,
    thick,
    densely dashed,
    mark=o,
    mark options={draw=UROOrange,fill=white}
]
coordinates {
    (0.1,10.627383)
    (0.2,21.526396)
    (0.3,819.683765)
    (0.4,2087.835745)
    (0.5,2661.427335)
    (0.6,2811.357840)
    (0.7,2780.837766)
    (0.8,2677.559970)
    (0.9,2508.554067)
    (1.0,2376.694746)
    (1.1,2223.178209)
    (1.2,2098.905108)
};
\addlegendentry{URO}

\end{axis}
\end{tikzpicture}
\caption{Rate-weighted average task delay versus total arrival rate
$\Lambda_{\mathrm{sum}} = \sum_k \lambda_k$ for the three-task MEC
system. The PFQN-based policy is shown with both its analytical
prediction and simulated performance, and is compared against
random routing (URO), proportional-to-$\mu$ routing (PO),
ITAGS, and EFO.}
\label{fig:totalrate-delay}
\end{figure}
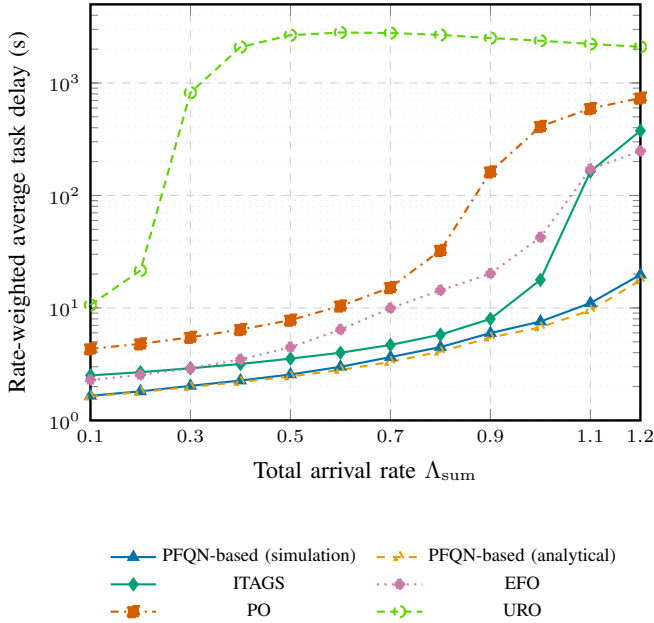

Figure~\ref{fig:totalrate-delay} reports the rate-weighted average task
delay as a function of the total arrival rate
$\Lambda_{\mathrm{sum}}$ in the mixed-traffic setting, where all
three task types are active simultaneously. For the PFQN-based policy,
we show both the analytically predicted and the simulated delay. The
analytical curve tracks the simulation curve closely over most of the
tested load range and captures the correct growth trend as the system
approaches heavy load. At the same time, the optimized policy
consistently outperforms all baseline methods, even dynamic ones, over the displayed
load sweep.

Table~\ref{tab:server-utilization} reports the server
utilizations under the compared policies. The results show that the
main bottlenecks under URO are the slower user-device queues, which
approach saturation very early. PO distributes the load more evenly
but its first three queues become nearly saturated as the arrival rate
approaches $0.90$. Under ITAGS and EFO, the edge server $S_3$ becomes
the dominant bottleneck. In contrast, the PFQN-based optimized policy
shifts work away from the most vulnerable bottlenecks and uses the
edge and cloud resources more effectively. At
$\Lambda_{\mathrm{sum}}=1.20$, its utilization vector is
$(0.810,0.838,0.916,0.928)$, while the other methods have at least one
nearly saturated queue. This is matched with the significant delay reductions
observed in Fig.~\ref{fig:totalrate-delay}.

\begin{table*}[t]
\centering
\caption{Server utilizations $\rho_n$ versus total arrival rate $\Lambda_{\mathrm{sum}}$ under different routing policies.}
\label{tab:server-utilization}
\setlength{\tabcolsep}{2pt}
\begin{tabular}{|c|cccc|cccc|cccc|cccc|cccc|}
\toprule
\multirow{2}{*}{$\Lambda_{\mathrm{sum}}$} &
\multicolumn{4}{c|}{URO} &
\multicolumn{4}{c|}{PO} &
\multicolumn{4}{c|}{ITAGS} &
\multicolumn{4}{c|}{EFO} &
\multicolumn{4}{c|}{PFQN-based} \\
& $\rho_1$ & $\rho_2$ & $\rho_3$ & $\rho_4$ &
  $\rho_1$ & $\rho_2$ & $\rho_3$ & $\rho_4$ &
  $\rho_1$ & $\rho_2$ & $\rho_3$ & $\rho_4$ &
  $\rho_1$ & $\rho_2$ & $\rho_3$ & $\rho_4$ &
  $\rho_1$ & $\rho_2$ & $\rho_3$ & $\rho_4$ \\
\midrule
0.10 & 0.384 & 0.328 & 0.062 & 0.066 & 0.115 & 0.116 & 0.115 & 0.092 & 0.000 & 0.000 & 0.191 & 0.063 & 0.000 & 0.000 & 0.137 & 0.074 & 0.001 & 0.001 & 0.046 & 0.122 \\
0.20 & 0.761 & 0.659 & 0.124 & 0.132 & 0.230 & 0.228 & 0.230 & 0.182 & 0.000 & 0.004 & 0.363 & 0.133 & 0.000 & 0.000 & 0.277 & 0.145 & 0.000 & 0.000 & 0.122 & 0.229 \\
0.30 & 0.998 & 0.945 & 0.180 & 0.188 & 0.347 & 0.346 & 0.345 & 0.274 & 0.002 & 0.017 & 0.513 & 0.213 & 0.000 & 0.000 & 0.420 & 0.216 & 0.000 & 0.000 & 0.206 & 0.312 \\
0.40 & 0.999 & 0.999 & 0.213 & 0.235 & 0.460 & 0.463 & 0.457 & 0.366 & 0.009 & 0.045 & 0.644 & 0.299 & 0.000 & 0.000 & 0.568 & 0.284 & 0.000 & 0.000 & 0.291 & 0.382 \\
0.50 & 0.999 & 0.999 & 0.242 & 0.276 & 0.572 & 0.573 & 0.572 & 0.459 & 0.029 & 0.096 & 0.746 & 0.398 & 0.001 & 0.005 & 0.711 & 0.356 & 0.000 & 0.000 & 0.393 & 0.455 \\
0.60 & 0.999 & 1.000 & 0.271 & 0.316 & 0.689 & 0.692 & 0.693 & 0.551 & 0.073 & 0.181 & 0.835 & 0.503 & 0.009 & 0.033 & 0.843 & 0.430 & 0.000 & 0.000 & 0.492 & 0.534 \\
0.70 & 1.000 & 1.000 & 0.299 & 0.357 & 0.802 & 0.797 & 0.801 & 0.640 & 0.160 & 0.303 & 0.897 & 0.611 & 0.071 & 0.161 & 0.943 & 0.510 & 0.000 & 0.000 & 0.587 & 0.619 \\
0.80 & 1.000 & 1.000 & 0.326 & 0.397 & 0.910 & 0.908 & 0.917 & 0.735 & 0.306 & 0.463 & 0.943 & 0.724 & 0.250 & 0.413 & 0.987 & 0.593 & 0.000 & 0.000 & 0.680 & 0.696 \\
0.90 & 1.000 & 1.000 & 0.353 & 0.438 & 0.977 & 0.979 & 0.993 & 0.801 & 0.513 & 0.650 & 0.974 & 0.831 & 0.545 & 0.699 & 0.998 & 0.676 & 0.000 & 0.000 & 0.767 & 0.776 \\
1.00 & 1.000 & 1.000 & 0.377 & 0.476 & 0.993 & 0.993 & 0.997 & 0.853 & 0.803 & 0.869 & 0.993 & 0.940 & 0.890 & 0.938 & 0.999 & 0.755 & 0.401 & 0.438 & 0.757 & 0.833 \\
1.10 & 1.000 & 1.000 & 0.404 & 0.518 & 0.996 & 0.996 & 0.999 & 0.902 & 0.991 & 0.994 & 1.000 & 0.997 & 0.983 & 0.988 & 0.999 & 0.833 & 0.649 & 0.671 & 0.844 & 0.866 \\
1.20 & 1.000 & 1.000 & 0.426 & 0.554 & 0.997 & 0.997 & 0.999 & 0.940 & 0.994 & 0.996 & 1.000 & 0.998 & 0.987 & 0.991 & 1.000 & 0.910 & 0.810 & 0.838 & 0.916 & 0.928 \\
\bottomrule
\end{tabular}
\end{table*}

To quantify the gain of the optimized policy relative to the
baselines, we report the percentage delay reduction defined by
\begin{equation}
\Delta_{\mathrm{alg}}(\Lambda_{\mathrm{sum}})
=
100\,
\frac{
T_{\mathrm{alg}}(\Lambda_{\mathrm{sum}})
-
T_{\mathrm{PFQN}}(\Lambda_{\mathrm{sum}})
}{
T_{\mathrm{alg}}(\Lambda_{\mathrm{sum}})
},
\label{eq:delay_reduction_metric}
\end{equation}
where $T_{\mathrm{PFQN}}(\Lambda_{\mathrm{sum}})$ denotes the rate-weighted average task delay of the optimized
PFQN-based policy, and
$T_{\mathrm{alg}}(\Lambda_{\mathrm{sum}})$ denotes the corresponding delay of the baseline method, all obtained by simulation.

\begin{table}[t]
\centering
\caption{Percentage delay reduction of the optimized PFQN-based policy
relative to the baseline methods at representative total arrival
rates.}
\label{tab:delay-reduction}
\setlength{\tabcolsep}{4.5pt}
\begin{tabular}{c c c c c}
\toprule
$\Lambda_{\mathrm{sum}}$
& vs. URO
& vs. PO
& vs. ITAGS
& vs. EFO \\
\midrule
$0.10$ & $84.41\%$ & $61.69\%$ & $34.07\%$ & $27.51\%$ \\
$0.20$ & $91.58\%$ & $62.18\%$ & $32.45\%$ & $28.73\%$ \\
$0.50$ & $99.90\%$ & $67.12\%$ & $27.67\%$ & $42.75\%$ \\
$0.70$ & $99.87\%$ & $75.92\%$ & $21.95\%$ & $63.21\%$ \\
$0.90$ & $99.76\%$ & $96.31\%$ & $25.19\%$ & $70.38\%$ \\
$1.20$ & $99.06\%$ & $97.31\%$ & $94.76\%$ & $92.05\%$ \\
\bottomrule
\end{tabular}
\end{table}

As Table~\ref{tab:delay-reduction} shows, the PFQN-based policy
achieves a lower delay than every baseline at all reported
arrival rates. Its advantage over URO and PO becomes particularly
large as their queues approach saturation. The improvement over EFO
also increases substantially in the medium-to-heavy load regime.
Although the improvement over ITAGS is more moderate at lower arrival
rates, it increases sharply at $\Lambda_{\mathrm{sum}}=1.20$.

For completeness, all available finite-horizon simulation results are
included in Table~\ref{tab:delay-reduction}. Some high-load entries
correspond to baseline methods operating beyond their
maximum stable throughput.

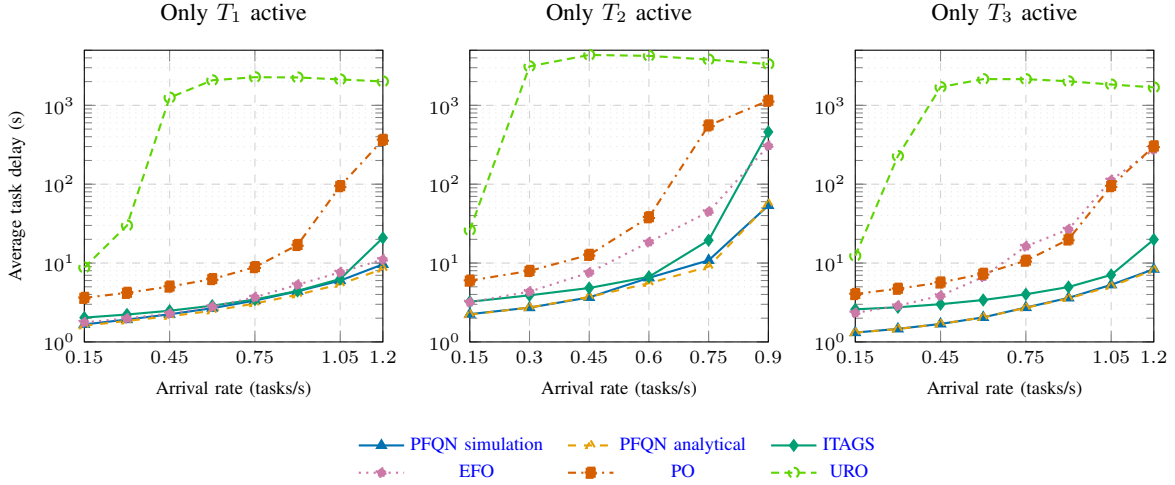
\begin{figure*}[t]
\centering
\begin{tikzpicture}
\begin{groupplot}[
    group style={
        group size=3 by 1,
        horizontal sep=1.15cm
    },
    width=0.305\textwidth,
    height=0.30\textwidth,
    ymode=log,
    log basis y=10,
    ymin=1,
    ymax=5000,
    grid=both,
    major grid style={dashed,gray!35},
    minor grid style={dotted,gray!20},
    tick label style={font=\scriptsize},
    label style={font=\scriptsize},
    title style={font=\small},
    every axis plot/.append style={mark size=1.9pt,line width=0.8pt},
    xlabel={Arrival rate (tasks/s)}
]

\nextgroupplot[
    title={Only $T_1$ active},
    ylabel={Average task delay (s)},
    xmin=0.15,
    xmax=1.20,
    xtick={0.15,0.45,0.75,1.05,1.20},
    legend to name=singleclasslegend,
    legend columns=3,
    legend style={
        draw=none,
        fill=none,
        font=\scriptsize,
        /tikz/every even column/.append style={column sep=0.7em}
    }
]

\addplot[
    color=PFQNBlue,
    thick,
    solid,
    mark=triangle*,
    mark options={draw=PFQNBlue,fill=PFQNBlue}
]
coordinates {
    (0.15,1.667636)
    (0.30,1.920295)
    (0.45,2.240275)
    (0.60,2.677574)
    (0.75,3.343585)
    (0.90,4.385771)
    (1.05,5.995337)
    (1.20,9.630012)
};
\addlegendentry{PFQN simulation}

\addplot[
    color=PFQNCyan,
    thick,
    dashed,
    mark=triangle,
    mark options={draw=PFQNCyan,fill=white}
]
coordinates {
    (0.15,1.604155)
    (0.30,1.847072)
    (0.45,2.116631)
    (0.60,2.480915)
    (0.75,3.068185)
    (0.90,3.962681)
    (1.05,5.393788)
    (1.20,8.434868)
};
\addlegendentry{PFQN analytical}

\addplot[
    color=ITAGSGreen,
    thick,
    solid,
    mark=diamond*,
    mark options={draw=ITAGSGreen,fill=ITAGSGreen}
]
coordinates {
    (0.15,2.034048)
    (0.30,2.225255)
    (0.45,2.485513)
    (0.60,2.879020)
    (0.75,3.452633)
    (0.90,4.436065)
    (1.05,6.292125)
    (1.20,20.853486)
};
\addlegendentry{ITAGS}

\addplot[
    color=EFOPurple,
    thick,
    dotted,
    mark=pentagon*,
    mark options={draw=EFOPurple,fill=EFOPurple}
]
coordinates {
    (0.15,1.770410)
    (0.30,1.986628)
    (0.45,2.285228)
    (0.60,2.792049)
    (0.75,3.687168)
    (0.90,5.312696)
    (1.05,7.641066)
    (1.20,11.094366)
};
\addlegendentry{EFO}

\addplot[
    color=PORed,
    thick,
    dashdotted,
    mark=square*,
    mark options={draw=PORed,fill=PORed}
]
coordinates {
    (0.15,3.621008)
    (0.30,4.192275)
    (0.45,5.034537)
    (0.60,6.265655)
    (0.75,8.889159)
    (0.90,16.958048)
    (1.05,94.491101)
    (1.20,364.664771)
};
\addlegendentry{PO}

\addplot[
    color=UROOrange,
    thick,
    densely dashed,
    mark=o,
    mark options={draw=UROOrange,fill=white}
]
coordinates {
    (0.15,8.783223)
    (0.30,29.889391)
    (0.45,1252.167318)
    (0.60,2078.135967)
    (0.75,2274.606494)
    (0.90,2257.204504)
    (1.05,2130.527736)
    (1.20,2008.648129)
};
\addlegendentry{URO}

\nextgroupplot[
    title={Only $T_2$ active},
    xmin=0.15,
    xmax=0.90,
    xtick={0.15,0.30,0.45,0.60,0.75,0.90}
]

\addplot[
    color=PFQNBlue,
    thick,
    solid,
    mark=triangle*,
    mark options={draw=PFQNBlue,fill=PFQNBlue}
]
coordinates {
    (0.15,2.250010)
    (0.30,2.722581)
    (0.45,3.669585)
    (0.60,6.493211)
    (0.75,10.816841)
    (0.90,53.735309)
};

\addplot[
    color=PFQNCyan,
    thick,
    dashed,
    mark=triangle,
    mark options={draw=PFQNCyan,fill=white}
]
coordinates {
    (0.15,2.220857)
    (0.30,2.728426)
    (0.45,3.724395)
    (0.60,5.540128)
    (0.75,8.995955)
    (0.90,55.976391)
};

\addplot[
    color=ITAGSGreen,
    thick,
    solid,
    mark=diamond*,
    mark options={draw=ITAGSGreen,fill=ITAGSGreen}
]
coordinates {
    (0.15,3.232218)
    (0.30,3.898689)
    (0.45,4.829573)
    (0.60,6.688738)
    (0.75,19.442155)
    (0.90,460.199090)
};

\addplot[
    color=EFOPurple,
    thick,
    dotted,
    mark=pentagon*,
    mark options={draw=EFOPurple,fill=EFOPurple}
]
coordinates {
    (0.15,3.180594)
    (0.30,4.340818)
    (0.45,7.562905)
    (0.60,18.388148)
    (0.75,44.710397)
    (0.90,305.942677)
};

\addplot[
    color=PORed,
    thick,
    dashdotted,
    mark=square*,
    mark options={draw=PORed,fill=PORed}
]
coordinates {
    (0.15,5.991055)
    (0.30,7.953499)
    (0.45,12.709719)
    (0.60,38.136040)
    (0.75,556.468529)
    (0.90,1143.384837)
};

\addplot[
    color=UROOrange,
    thick,
    densely dashed,
    mark=o,
    mark options={draw=UROOrange,fill=white}
]
coordinates {
    (0.15,26.148937)
    (0.30,3120.042419)
    (0.45,4353.139362)
    (0.60,4239.863631)
    (0.75,3803.515351)
    (0.90,3329.289176)
};

\nextgroupplot[
    title={Only $T_3$ active},
    xmin=0.15,
    xmax=1.20,
    xtick={0.15,0.45,0.75,1.05,1.20}
]

\addplot[
    color=PFQNBlue,
    thick,
    solid,
    mark=triangle*,
    mark options={draw=PFQNBlue,fill=PFQNBlue}
]
coordinates {
    (0.15,1.313969)
    (0.30,1.463397)
    (0.45,1.691850)
    (0.60,2.046699)
    (0.75,2.730935)
    (0.90,3.617165)
    (1.05,5.301457)
    (1.20,8.361103)
};

\addplot[
    color=PFQNCyan,
    thick,
    dashed,
    mark=triangle,
    mark options={draw=PFQNCyan,fill=white}
]
coordinates {
    (0.15,1.312095)
    (0.30,1.459020)
    (0.45,1.688812)
    (0.60,2.049853)
    (0.75,2.709112)
    (0.90,3.562855)
    (1.05,5.084444)
    (1.20,8.148928)
};

\addplot[
    color=ITAGSGreen,
    thick,
    solid,
    mark=diamond*,
    mark options={draw=ITAGSGreen,fill=ITAGSGreen}
]
coordinates {
    (0.15,2.588990)
    (0.30,2.749041)
    (0.45,3.009464)
    (0.60,3.399353)
    (0.75,4.018936)
    (0.90,4.968196)
    (1.05,7.075977)
    (1.20,19.873553)
};

\addplot[
    color=EFOPurple,
    thick,
    dotted,
    mark=pentagon*,
    mark options={draw=EFOPurple,fill=EFOPurple}
]
coordinates {
    (0.15,2.332126)
    (0.30,2.872042)
    (0.45,3.854425)
    (0.60,6.675761)
    (0.75,16.288064)
    (0.90,26.772826)
    (1.05,112.672770)
    (1.20,275.976940)
};

\addplot[
    color=PORed,
    thick,
    dashdotted,
    mark=square*,
    mark options={draw=PORed,fill=PORed}
]
coordinates {
    (0.15,4.052454)
    (0.30,4.726435)
    (0.45,5.670812)
    (0.60,7.312620)
    (0.75,10.722457)
    (0.90,19.919874)
    (1.05,94.604013)
    (1.20,302.293331)
};

\addplot[
    color=UROOrange,
    thick,
    densely dashed,
    mark=o,
    mark options={draw=UROOrange,fill=white}
]
coordinates {
    (0.15,12.343079)
    (0.30,227.506225)
    (0.45,1703.106875)
    (0.60,2150.610332)
    (0.75,2148.191468)
    (0.90,2018.838794)
    (1.05,1837.503170)
    (1.20,1688.267751)
};

\end{groupplot}

\node at ($(group c2r1.south)+(0,-1.55cm)$)
    {\ref{singleclasslegend}};

\end{tikzpicture}
\caption{Average task delay under single-class traffic scenarios:
(left) only type-$1$ tasks arrive; (middle) only type-$2$ tasks
arrive; (right) only type-$3$ tasks arrive. In all three cases, the
PFQN-based optimized policy achieves the lowest delay over the tested
load range.}
\label{fig:singleclass}
\end{figure*}

To further examine the robustness of the optimized policy with
respect to task-graph structure, we also consider three single-class
traffic scenarios in which only one task type is active at a time.
Figure~\ref{fig:singleclass} reports the resulting average delay when
the arrival stream consists solely of type-$1$, type-$2$, or type-$3$
tasks, respectively. In all three cases, the PFQN-based optimized
policy remains the best-performing method over the tested load range.
This shows that the gain of the proposed policy is not merely a
consequence of averaging across heterogeneous traffic classes, but
persists even when each task type is studied in isolation.

\subsection{Maximum Stable Throughput}

In queueing-network theory, the stability region is commonly defined
as the set of arrival-rate vectors for which the network queues can be
stabilized, and the maximum achievable throughput is characterized by
the boundary of this region~\cite{Tassiulas1992,Neely2010}. Queue
stability has also been explicitly considered in computation-offloading
systems with stochastic task arrivals~\cite{Bi2021LyDROO}. Motivated by
these definitions, we use the maximum stable throughput to
describe the stability boundary observed in our simulations.

Since the stability boundary is estimated through finite-duration
simulations with finite search resolution, it is reported as
$
    \left[
    \Lambda_{\mathrm{stable}},
    \Lambda_{\mathrm{unstable}}
    \right),
$
where $\Lambda_{\mathrm{stable}}$ is the largest tested arrival rate
that satisfies the empirical stability criterion and
$\Lambda_{\mathrm{unstable}}$ is the first tested arrival rate that
does not satisfy it. Thus, the reported interval provides an empirical
estimate of the stability boundary rather than an analytical guarantee.

Table~\ref{tab:stable-throughput} reports the maximum stable
throughput intervals under mixed traffic and the three single-class
traffic scenarios. The final column reports the improvement of the
PFQN-based policy over the best baseline, calculated using the
largest empirically stable rates.

\begin{table*}[t]
\centering
\caption{Maximum stable throughput intervals in tasks/s.
Each entry is reported as
$[\Lambda_{\mathrm{stable}},\Lambda_{\mathrm{unstable}})$, where
$\Lambda_{\mathrm{stable}}$ is the largest tested empirically stable
arrival rate and $\Lambda_{\mathrm{unstable}}$ is the first tested
unstable arrival rate.}
\label{tab:stable-throughput}
\scriptsize
\setlength{\tabcolsep}{3.2pt}
\renewcommand{\arraystretch}{1.15}
\begin{tabular}{c c c c c c c}
\toprule
Scenario
& PFQN-based
& URO
& PO
& ITAGS
& EFO
& PFQN gain \\
\midrule

Mixed
& $[1.240,1.243)$
& $[0.246,0.250)$
& $[0.843,0.846)$
& $[0.906,0.909)$
& $[0.746,0.750)$
& $36.90\%$ \\

Only $T_1$
& $[1.400,1.403)$
& $[0.318,0.321)$
& $[0.978,0.981)$
& $[1.087,1.090)$
& $[1.053,1.056)$
& $28.74\%$ \\

Only $T_2$
& $[0.884,0.887)$
& $[0.200,0.203)$
& $[0.637,0.640)$
& $[0.700,0.703)$
& $[0.559,0.562)$
& $26.34\%$ \\

Only $T_3$
& $[1.446,1.450)$
& $[0.281,0.284)$
& $[1.000,1.003)$
& $[1.037,1.040)$
& $[0.737,0.740)$
& $39.46\%$ \\

\bottomrule
\end{tabular}
\end{table*}

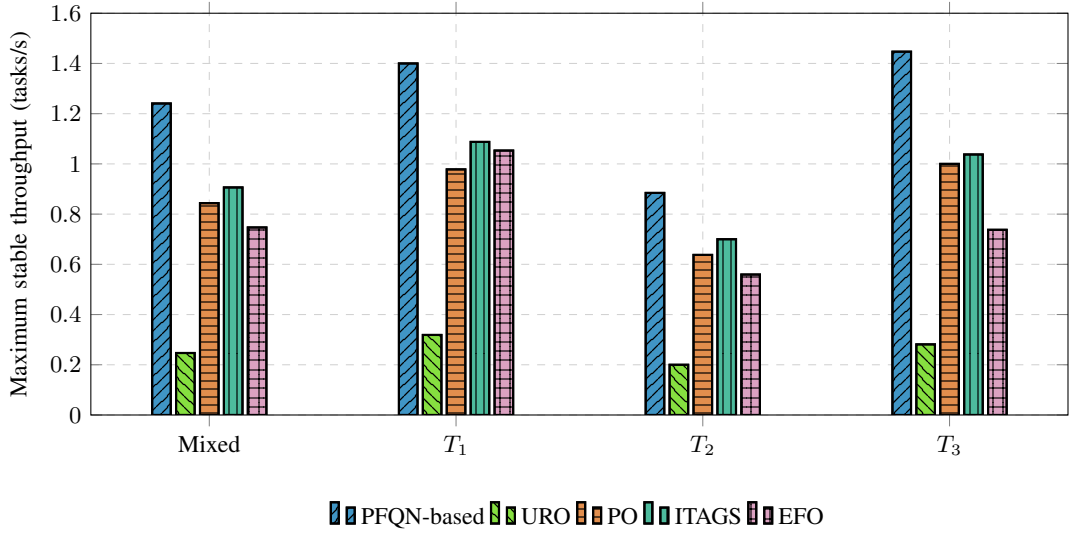
\begin{figure*}[t]
\centering
\begin{tikzpicture}
\begin{axis}[
    ybar,
    width=0.80\textwidth,
    height=0.32\textwidth,
    bar width=7pt,
    ymin=0,
    ymax=1.6,
    ytick distance=0.2,
    ylabel={Maximum stable throughput (tasks/s)},
    symbolic x coords={Mixed,T1,T2,T3},
    xtick=data,
    xticklabels={Mixed,$T_1$,$T_2$,$T_3$},
    enlarge x limits=0.16,
    grid=major,
    major grid style={dashed,gray!35},
    tick label style={font=\small},
    label style={font=\small},
    legend style={
        at={(0.5,-0.20)},
        anchor=north,
        legend columns=5,
        draw=none,
        fill=none,
        font=\small
    }
]

\addplot[
    draw=black,
    thick,
    preaction={fill=PFQNBlue!75},
    pattern={
        Lines[
            angle=45,
            distance=3pt,
            line width=0.45pt
        ]
    },
    pattern color=black
]
coordinates {
    (Mixed,1.240625)
    (T1,1.400000)
    (T2,0.884375)
    (T3,1.446875)
};
\addlegendentry{PFQN-based}

\addplot[
    draw=black,
    thick,
    preaction={fill=UROOrange!75},
    pattern={
        Lines[
            angle=-45,
            distance=3pt,
            line width=0.45pt
        ]
    },
    pattern color=black
]
coordinates {
    (Mixed,0.246875)
    (T1,0.318750)
    (T2,0.200000)
    (T3,0.281250)
};
\addlegendentry{URO}

\addplot[
    draw=black,
    thick,
    preaction={fill=PORed!70},
    pattern={
        Lines[
            angle=0,
            distance=3pt,
            line width=0.45pt
        ]
    },
    pattern color=black
]
coordinates {
    (Mixed,0.843750)
    (T1,0.978125)
    (T2,0.637500)
    (T3,1.000000)
};
\addlegendentry{PO}

\addplot[
    draw=black,
    thick,
    preaction={fill=ITAGSGreen!70},
    pattern={
        Lines[
            angle=90,
            distance=3pt,
            line width=0.45pt
        ]
    },
    pattern color=black
]
coordinates {
    (Mixed,0.906250)
    (T1,1.087500)
    (T2,0.700000)
    (T3,1.037500)
};
\addlegendentry{ITAGS}

\addplot[
    draw=black,
    thick,
    preaction={fill=EFOPurple!70},
    pattern={
        Hatch[
            distance=3.5pt,
            line width=0.40pt
        ]
    },
    pattern color=black
]
coordinates {
    (Mixed,0.746875)
    (T1,1.053125)
    (T2,0.559375)
    (T3,0.737500)
};
\addlegendentry{EFO}

\end{axis}
\end{tikzpicture}

\caption{Maximum stable throughput under mixed traffic and
the three single-class traffic scenarios. Each bar shows the largest
tested arrival rate satisfying the empirical stability criterion for
the corresponding routing policy. The complete stability-boundary
intervals are reported in Table~\ref{tab:stable-throughput}. }
\label{fig:stable-throughput-comparison}
\end{figure*}

As shown in Table~\ref{tab:stable-throughput} and
Fig.~\ref{fig:stable-throughput-comparison}, the PFQN-based policy
achieves the highest stable maximum throughput in all four
traffic scenarios. In mixed traffic, the highest stable
rate of the PFQN-based policy is $1.240$ tasks/s, compared to
$0.906$ tasks / s for ITAGS, the best baseline. This corresponds
to an improvement of $36.90\%$. In the $T_1$, $T_2$, and $T_3$
single-class scenarios, the corresponding improvements over the
best baseline are $28.74\%$, $26.34\%$, and $39.46\%$,
respectively. These results indicate that the PFQN-based policy not
only reduces average task delay within the stable operating region,
but also enables the MEC system to accommodate higher aggregate
task-arrival rates than the considered baseline policies.

\section{Conclusion}
\label{sec:conclusion}
In this paper, we developed an analytical framework for static
computation offloading in multi-tier MEC systems with
tree-structured task graphs. To capture the correlated and batched
internal traffic generated by dependent subtasks, we proposed a
product-form queueing-network approximation based on Poissonization
and a local congestion-correction term. This yields tractable
closed-form expressions for effective utilizations and waiting times.

Building on these queueing parameters, we derived a recursive delay
evaluation algorithm for tree-structured tasks and used it to optimize
the static routing probabilities through a differentiable formulation
based on softmax parameterization, log-sum-exp smoothing, and a
stability barrier. Numerical results demonstrated that the proposed approximation accurately predicts average task delay over a broad family of stable routing policies. The delay-optimized static policy also achieved the lowest delay within the stable operating region and increased the maximum stable throughput by 26.34\%–39.46\% relative to the best baseline across the evaluated traffic scenarios.

\appendices
\section{Derivation of the Congestion Correction Term}
\label{app:correction}

In this appendix, we derive the correction term used in \eqref{eq:C_n}--\eqref{eq:alpha_closed}. We consider a single server $S_n$, so for notational simplicity we omit the index $n$.

Suppose that the server receives several types of batches. A batch of type $(k,m)$ arrives with initiation rate $\gamma_{km}(R)$ and contains exactly $B_{km}$ customers. We compare two systems with the same exponential service rate $\mu$ and FCFS discipline.

\subsection{Original Batch-Arrived Queue}

In the original queue, every initiated batch of type $(k,m)$ contributes $B_{km}$ customers. Let $T_p$ denote the average response time seen by the \emph{first} customer of a batch. Then the second customer in that batch sees one extra customer ahead of it and has average response time $T_p+\frac{1}{\mu}$, the third customer has average response time $T_p+\frac{2}{\mu}$, and so on, up to the last customer whose average response time is
\[
T_p+\frac{B_{km}-1}{\mu}.
\]
Hence, by Little's law, the total average number of customers contributed by type $(k,m)$ batches is
\begin{multline}
\gamma_{km}(R)\sum_{r=0}^{B_{km}-1}\left(T_p+\frac{r}{\mu}\right)
=
\\
\gamma_{km}(R)\left(B_{km}T_p+\frac{B_{km}(B_{km}-1)}{2\mu}\right).
\label{eq:Np_ap}
\end{multline}
Summing over all batch types, the mean number of customers in the original queue is
\begin{equation}
N_p
=
\sum_{k=1}^{K}\sum_{m=1}^{M_k}
\gamma_{km}(R)
\left(
B_{km}T_p+\frac{B_{km}(B_{km}-1)}{2\mu}
\right).
\label{eq:Np_app}
\end{equation}

\subsection{Approximate Dual Queue}

In the approximate dual queue, each batch of size $B_{km}$ is replaced by a Poisson stream of individual customers with rate $B_{km}\gamma_{km}(R)$. In addition, we insert an independent Poisson stream of redundant customers with rate $\alpha(R)$. Let $T_d$ denote the mean response time in this approximate M/M/1 queue. Then the mean number of customers in the approximate dual queue is
\begin{equation}
N_d
=
\left(
\sum_{k=1}^{K}\sum_{m=1}^{M_k} B_{km}\gamma_{km}(R)
+\alpha(R)
\right)T_d.
\label{eq:Nd_app}
\end{equation}

\subsection{Matching the Two Queues}

The correction rate $\alpha(R)$ is chosen so that the two queues have the same average number of customers:
\begin{equation}
N_p=N_d.
\label{eq:match_app}
\end{equation}
Since both queues are single-server FCFS queues with the same exponential service rate, equality of the average number of customers implies equality of the response time seen by a Poisson arrival. Therefore,
\[
T_p=T_d.
\]
Using \eqref{eq:Np_app}, \eqref{eq:Nd_app}, and $T_p=T_d$, we obtain
\[
\alpha(R)T_d
=
\sum_{k=1}^{K}\sum_{m=1}^{M_k}
\gamma_{km}(R)\frac{B_{km}(B_{km}-1)}{2\mu}.
\]
Define
\begin{equation}
C(R)
=
\sum_{k=1}^{K}\sum_{m=1}^{M_k}
\frac{B_{km}(B_{km}-1)}{2}\gamma_{km}(R).
\label{eq:C_app}
\end{equation}
Then
\[
\alpha(R)T_d=\frac{C(R)}{\mu}.
\]
Since the approximating queue is M/M/1 with utilization
\[
\rho(R)=\frac{\Lambda(R)+\alpha(R)}{\mu},
\]
its mean response time is
\[
T_d=\frac{1}{\mu(1-\rho(R))}.
\]
Substituting this into the previous equation yields
\[
\alpha(R)=\bigl(1-\rho(R)\bigr)C(R).
\]
Therefore,
\begin{equation}
\alpha(R)=\bigl(1-\rho(R)\bigr)C(R),
\qquad
\rho(R)=\frac{\Lambda(R)+\alpha(R)}{\mu},
\label{eq:alpha_app_fixed}
\end{equation}
which is exactly the relation used in Section~\ref{sec:qn_model}. Solving \eqref{eq:alpha_app_fixed} gives
\[
\rho(R)=\frac{\Lambda(R)+C(R)}{\mu+C(R)},
\qquad
\alpha(R)=\frac{C(R)\bigl(\mu-\Lambda(R)\bigr)}{\mu+C(R)}.
\]

Applying this derivation separately to each server $S_n$ yields \eqref{eq:rho_closed} and \eqref{eq:alpha_closed}.


\begin{thebibliography}{99}

\bibitem{MaoMECsurvey}
Y.~Mao, C.~You, J.~Zhang, K.~Huang, and K.~B.~Letaief,
``A survey on mobile edge computing: The communication perspective,''
\emph{IEEE Commun. Surveys Tuts.}, vol.~19, no.~4, pp.~2322--2358, 2017.

\bibitem{DinhMCCsurvey}
H.~T.~Dinh, C.~Lee, D.~Niyato, and P.~Wang,
``A survey of mobile cloud computing: Architecture, applications, and approaches,''
\emph{Wireless Commun. Mobile Comput.}, vol.~13, no.~18, pp.~1587--1611, Dec.~2013.

\bibitem{SatyanarayananEdge}
M.~Satyanarayanan,
``The emergence of edge computing,''
\emph{Computer}, vol.~50, no.~1, pp.~30--39, Jan.~2017.

\bibitem{ChenMultiUserOffload16}
X.~Chen, L.~Jiao, W.~Li, and X.~Fu,
``Efficient multi-user computation offloading for mobile-edge cloud computing,''
\emph{IEEE/ACM Trans. Netw.}, vol.~24, no.~5, pp.~2795--2808, Oct.~2016.

\bibitem{QECO2025}
I. Rahmaty, H. Shah-Mansouri, and A. Movaghar,
``QECO: A QoE-oriented computation offloading algorithm based on
deep reinforcement learning for mobile edge computing,''
\emph{IEEE Trans. Netw. Sci. Eng.}, vol. 12, no. 4,
pp. 3118--3130, Jul.--Aug. 2025,
doi: 10.1109/TNSE.2025.3556809.

\bibitem{DependentTaskGraphDRL2025}
R. Guo, L. Zhou, L. Li, Y. Song, and X. Xie,
``Dependent task graph offloading model based on deep reinforcement
learning in mobile edge computing,''
\emph{Electronics}, vol. 14, no. 16, Art. no. 3184, Aug. 2025,
doi: 10.3390/electronics14163184.

\bibitem{Zhai2025Joint}
L. Zhai, Z. Lu, J. Sun, and X. Li,
``Joint task offloading and computing resource allocation with DQN
for task-dependency in multi-access edge computing,''
\emph{Computer Networks}, vol. 263, Art. no. 111222, 2025,
doi: 10.1016/j.comnet.2025.111222.

\bibitem{Sardellitti2015}
S.~Sardellitti, G.~Scutari, and S.~Barbarossa,
``Joint optimization of radio and computational resources for multicell mobile-edge computing,''
\emph{IEEE Trans. Signal Inf. Process. Netw.}, vol.~1, no.~2, pp.~89--103, Jun.~2015.

\bibitem{TaskGraphMEC}
J.~Yan, S.~Bi, and Y.-J.~A.~Zhang,
``Offloading and resource allocation with general task graph in mobile edge computing: A deep reinforcement learning approach,''
\emph{IEEE Trans. Wireless Commun.}, vol.~19, no.~8, pp.~5404--5419, Aug.~2020.

\bibitem{CormenAlgorithms09}
T.~H.~Cormen, C.~E.~Leiserson, R.~L.~Rivest, and C.~Stein,
\emph{Introduction to Algorithms}, 3rd~ed.
Cambridge, MA, USA: MIT Press, 2009.

\bibitem{LiIBDASH22}
X. Li, M. Abdallah, S. Suryavansh, M. Chiang, K. T. Kim,
and S. Bagchi,
``DAG-based task orchestration for edge computing,''
in \emph{Proc. 41st IEEE Int. Symp. Reliable Distributed Systems
(SRDS)}, 2022, pp. 23--34,
doi: 10.1109/SRDS55811.2022.00013.

\bibitem{XueQueueMEC}
J.~Xue, Z.~Wang, Y.~Zhang, and L.~Wang,
``Task Allocation Optimization Scheme Based on Queuing Theory for Mobile Edge Computing in 5G Heterogeneous Networks.''
\emph{Mobile Information Systems}, vol.~2020, Art.~ID~1501403, May~2020, doi:~10.1155/2020/1501403.

\bibitem{KatayamaQueueSensors22}
Y.~Katayama and T.~Tachibana,
``Optimal Task Allocation Algorithm Based on Queueing Theory for Future Internet Application in Mobile Edge Computing Platform,''
\emph{Sensors}, vol.~22, no.~13, Art.~no.~4825, 2022, doi:~10.3390/s22134825.

\bibitem{BairagiQueueLengthCCNC24}
R.-H.~Hwang, Y.-C.~Lai, and Y.-D.~Lin,
``Queue-Length-Based Offloading for Delay Sensitive Applications in Federated Cloud-Edge-Fog Systems,''
in \emph{Proc. IEEE CCNC}, 2024, pp.~406--411, doi:~10.1109/CCNC51664.2024.10454697.

\bibitem{XieMarkov23}
Z. Xie, X. Zhao, D. Han, L. Gao, Z. Jiang, J. Zhu, X. She,
and P. Chen,
``A Markovian queueing model for end-to-end delay analysis in
computation offloading system,''
\emph{IEEE Commun. Lett.}, vol. 27, no. 10,
pp. 2687--2691, Oct. 2023,
doi: 10.1109/LCOMM.2023.3307876.


\bibitem{DongSurvey2024}
S.~Dong \emph{et~al.},
``Task offloading strategies for mobile edge computing: A survey,''
\emph{Computer Networks}, vol.~254, Art.~no.~110791, 2024, doi:~10.1016/j.comnet.2024.110791.

\bibitem{WangRobustOffloadingTMC2023}
H.~Wang, H.~Xu, H.~Huang, M.~Chen, and S.~Chen,
``Robust task offloading in dynamic edge computing,''
\emph{IEEE Trans. Mobile Comput.}, vol.~22, no.~1, pp.~500--514, Jan.~2023.

\bibitem{QuRobustSchedulingTMC2022}
Y.~Qu, H.~Dai, F.~Wu, D.~Lu, C.~Dong, S.~Tang, and G.~Chen,
``Robust offloading scheduling for mobile edge computing,''
\emph{IEEE Trans. Mobile Comput.}, vol.~21, no.~7, pp.~2581--2595, Jul.~2022.

\bibitem{LiResilienceJSAC2023}
S.~Li, C.~Li, Y.~Huang, B.~A.~Jalaian, Y.~T.~Hou, and W.~Lou,
``Enhancing resilience in mobile edge computing under processing uncertainty,''
\emph{IEEE J. Sel. Areas Commun.}, vol.~41, no.~3, pp.~659--674, Mar.~2023.

\bibitem{JiaoJCC24}
S.~Jiao, H.~Wang, and J.~Luo,
``SRA-E-ABCO: Terminal task offloading for cloud-edge-end environments,''
\emph{J. Cloud Computing}, vol.~13, no.~1, Art.~no.~58, 2024, doi:~10.1186/s13677-024-00622-y.

\bibitem{SundarLiang2018}
S.~Sundar and B.~Liang, ``Offloading Dependent Tasks with Communication Delay and Deadline Constraint,'' in \emph{Proc. IEEE INFOCOM}, Apr. 2018, pp.~37--45, doi:~10.1109/INFOCOM.2018.8486305.

\bibitem{ShuZhaoHanMinDuan2019}
C.~Shu, Z.~Zhao, Y.~Han, G.~Min, and H.~Duan, ``Multi-user offloading for edge computing networks: A
dependency-aware and latency-optimal approach,''
\emph{IEEE Internet of Things Journal}, vol. 7, no. 3,
pp. 1678--1689, Mar. 2020, doi: 10.1109/JIOT.2019.2943373.


\bibitem{CaoDependentOffloading2023}
Z. Cao, X. Deng, S. Yue, P. Jiang, J. Ren, and J. Gui,
``Dependent task offloading in edge computing using GNN and deep
reinforcement learning,''
\emph{IEEE Internet Things J.}, vol. 11, no. 12,
pp. 21632--21646, Jun. 2024,
doi: 10.1109/JIOT.2024.3374969.


\bibitem{AleTCCN21}
L. Ale, N. Zhang, X. Fang, X. Chen, S. Wu, and L. Li,
``Delay-aware and energy-efficient computation offloading in
mobile-edge computing using deep reinforcement learning,''
\emph{IEEE Transactions on Cognitive Communications and Networking},
vol. 7, no. 3, pp. 881--892, Sep. 2021,
doi: 10.1109/TCCN.2021.3066619.


\bibitem{BCMP75}
F.~Baskett, K.~M.~Chandy, R.~R.~Muntz, and F.~G.~Palacios,
``Open, closed, and mixed networks of queues with different classes of customers,''
\emph{J. ACM}, vol.~22, no.~2, pp.~248--260, 1975.

\bibitem{Kelly79}
F.~P.~Kelly,
\emph{Reversibility and Stochastic Networks}.
Chichester, U.K.: Wiley, 1979.

\bibitem{Serfozo99}
R.~Serfozo,
\emph{Introduction to Stochastic Networks}.
New York, USA: Springer, 1999.

\bibitem{Gelenbe93}
E. Gelenbe and G. Pujolle,
\emph{Introduction to Queueing Networks}, 2nd ed.
Chichester, U.K.: Wiley, 1998.

\bibitem{RahnamaniaAshtianiCorrelatedArrivals}
M.~Rahnamania and F.~Ashtiani,
``A new analytical approach for delay analysis in the presence of correlated arrivals,''
in \emph{Proc. Iran Workshop Commun. Inf. Theory (IWCIT)}, 2024, pp.~1--6, doi:~10.1109/IWCIT62550.2024.10553217.

\bibitem{RahnamaniaAshtianiTriggeredRequests}
M. Rahnamania and F. Ashtiani, ``Delay Analysis for Triggered Requests: M/G/1 Modeling Approach,'' \emph{IEEE INFOCOM 2026} - IEEE Conference on Computer Communications, Tokyo, Japan, 2026, pp. 1-10, doi: 10.1109/INFOCOM59046.2026.11571174.


\bibitem{BoydConvex}
S.~Boyd and L.~Vandenberghe,
\emph{Convex Optimization}.
Cambridge, U.K.: Cambridge Univ. Press, 2004.

\bibitem{BertsekasNonlinear}
D.~P.~Bertsekas,
\emph{Nonlinear Programming}, 2nd~ed.
Belmont, MA, USA: Athena Scientific, 1999.

\bibitem{KleinrockQueueing75}
L.~Kleinrock,
\emph{Queueing Systems, Volume I: Theory}.
New York, NY, USA: Wiley, 1975.

\bibitem{ZhangWirelessService23}
Y. Zhang, Y. Jiang, and S. Fu,
``Service modeling and delay analysis of packet delivery over a
wireless link,''
in \emph{Proc. IEEE 28th Int. Workshop Computer Aided Modeling
and Design of Communication Links and Networks (CAMAD)},
Edinburgh, U.K., Nov. 2023, pp. 296--301,
doi: 10.1109/CAMAD59638.2023.10478403.

\bibitem{Ahmed2019Entropy}
Z.~Ahmed, N.~Le Roux, M.~Norouzi, and D.~Schuurmans,
``Understanding the impact of entropy on policy optimization,''
in \emph{Proc. 36th Int. Conf. Mach. Learn. (ICML)},
vol.~97, pp.~151--160, 2019.

\bibitem{Mei2020Softmax}
J.~Mei, C.~Xiao, C.~Szepesv{\'a}ri, and D.~Schuurmans,
``On the global convergence rates of softmax policy gradient methods,''
in \emph{Proc. 37th Int. Conf. Mach. Learn. (ICML)},
vol.~119, pp.~6820--6829, 2020.

\bibitem{Tassiulas1992}
L.~Tassiulas and A.~Ephremides,
``Stability properties of constrained queueing systems and scheduling
policies for maximum throughput in multihop radio networks,''
\emph{IEEE Trans. Autom. Control},
vol.~37, no.~12, pp.~1936--1948, Dec.~1992,
doi: 10.1109/9.182479.

\bibitem{Neely2010}
M.~J. Neely,
\emph{Stochastic Network Optimization with Application to
Communication and Queueing Systems}.
San Rafael, CA, USA: Morgan \& Claypool, 2010,
doi: 10.2200/S00271ED1V01Y201006CNT007.

\bibitem{Bi2021LyDROO}
S.~Bi, L.~Huang, H.~Wang, and Y.-J. A.~Zhang,
``Lyapunov-guided deep reinforcement learning for stable online
computation offloading in mobile-edge computing networks,''
\emph{IEEE Trans. Wireless Commun.},
vol.~20, no.~11, pp.~7519--7537, Nov.~2021,
doi: 10.1109/TWC.2021.3085319.


\end{thebibliography}
\end{document}